\documentclass[12pt,preprint]{emulateapj}

\usepackage{epsf}
\usepackage{color}
\usepackage{amsmath}
\usepackage{apjfonts}

\newcommand{\hii}{\mbox{${\rm H}\,${\sc ii}}}
\newcommand{\sol}{$_{\odot}$}
\def\farcs{\hbox{$.\!\!^{\prime\prime}$}}
\def\arcs{\hbox{$^{\prime\prime}$}}
\def\arcm{\hbox{$^{\prime}$}}
\def\farcm{\hbox{$.\!\!^{\prime}$}}

\makeatletter

\newcommand{\Rmnum}[1]{\expandafter\@slowromancap\romannumeral #1@}
\makeatother

\shorttitle{\textit{Spitzer} Observations of NGC2467}
\shortauthors{Snider et al.}

\begin{document}
\slugcomment{Accepted for Publication in the Astrophysical Journal}
\title{\textit{Spitzer} Observations of the \hii~Region NGC 2467: An Analysis of Triggered Star Formation\altaffilmark{1}}

\author{\sc Keely D. Snider\altaffilmark{2}}
\affil{Department  of  Physics,  Arizona  State University,  Tempe, AZ  85287}
\affil{George P. and Cynthia Mitchell Institute for Fundamental Physics and Astronomy.  Department of Physics, Texas A\&M University, College Station, TX 77843}

\author{\sc  J.Jeff Hester, Steven J. Desch, Kevin R. Healy}
\affil{School of Earth and Space Exploration,  Arizona  State University,  Tempe, AZ  85287}

\and
 
\author{\sc John Bally}
\affil{Center for Astrophysics and Space Astronomy, University of Colorado, Boulder, CO 80309}

\altaffiltext{1}{Based on observations with the \textit{Spitzer Space Telescope}}
\altaffiltext{2}{Email: keely.snider@asu.edu}

\begin{abstract}
 
  We present new \textit{Spitzer Space Telescope} observations of the
  region NGC 2467, and use these observations to determine how the
  environment of an \hii~region affects the process of star formation.
  Our observations comprise IRAC (3.6, 4.5, 5.8, and 8.0 $\mu$m) and
  MIPS (24 $\mu$m) maps of the region, covering approximately 400
  square arcminutes.  The images show a region of ionized gas pushing
  out into the surrounding molecular cloud, powered by an O6V star and
  two clusters of massive stars in the region.  We have identified as
  candidate Young Stellar Objects (YSOs) 45 sources in NGC 2467 with
  infrared excesses in at least two mid-infrared colors.  We have
  constructed color-color diagrams of these sources and have
  quantified their spatial distribution within the region.  We find
  that the YSOs are not randomly distributed in NGC 2467; rather, over
  75\% of the sources are distributed at the edge of the \hii~region,
  along ionization fronts driven by the nearby massive stars.  The
  high fraction of YSOs in NGC 2467 that are found in proximity to gas
  that has been compressed by ionization fronts supports the
  hypothesis that a significant fraction of the star formation in NGC
  2467 is triggered by the massive stars and the expansion of the
  \hii~region.  At the current rate of star formation, we estimate at
  least 25-50\% of the total population of YSOs formed by this
  process.
\end{abstract}


\section{Introduction}   

The process of star formation is still not well understood, especially
in high-mass-star-forming regions.  One of the outstanding mysteries
is the degree to which the stars already formed in such regions,
especially the massive O or B stars, can affect the future formation
of stars in the region.  In some high-mass-star-forming regions there
is strong evidence for sequential star formation.  The
Scorpius-Centaurus (Sco-Cen) OB association is one extensively studied
example.  Stellar populations in the different subgroups of Sco-Cen
have measurably different ages that suggest a sequence of star
formation, with the oldest subgroup (Upper Centaurus-Lupus) located
between two younger subgroups (Lower Centaurus-Crux and Upper
Scorpius).  The implication is that the oldest subgroup formed in the
middle of a molecular cloud after some delay triggered the formation
of the two younger subgroups on either side of it (de Geus et al.\
1989; Preibisch \& Zinnecker 1999, 2001).

Other recent evidence for sequential star formation on smaller scales
has been observed in bright-rimmed clouds (BRCs) such as BRC 37 by
(Ikeda et al.\ 2008) and in the Orion OB association and Lac OB1
association (Lee et al.\ 2005; Lee $\&$ Chen 2007).  In these last two
associations new protostars are observed to line up between the
massive stars and the parent molecular clouds in the Orion and Lac OB
associations, with no young stars found embedded in the molecular
clouds far behind the ionization fronts, suggesting the stars form
only {\it after} passage of an ionization front launched by the
massive stars (Lee et al.\ 2005).  Thus evidence exists that massive
stars can affect or perhaps even cause the subsequent formation of
stars in a region.

An alternative interpretation is that even in high-mass-star-forming
regions all star formation is coeval (no delay between the formation
of high-mass stars and any other stars in the region).  In an extreme
form of this interpretation, massive stars can not modify subsequent
star formation because there is none; it has already happened.
Hillenbrand et al.\ (2008; White \& Hillenbrand 2004) claim that star
formation in \hii~regions environments is coeval and suggest that
estimates of ages for low-mass stars have been underestimated and
estimates of high-mass stars overestimated.  Hillenbrand et al.\
(2008) use isochronal fitting to estimate the ages of sources in many
star-forming regions and conclude there is no strong evidence for
moderate age spreads in either young star-forming regions or young
open clusters.

If we are to understand star formation generally, this difference of
interpretation must be resolved.  In their complete census of embedded
clusters (out to 2 kpc), Lada $\&$ Lada (2003) have shown that the
vast majority (70-90\%) of all (low-mass) stars formed in rich
clusters; roughly 50\% of all stars formed in clusters rich enough to
contain at least one O or B star capable of shaping its environment.
If a large fraction (say, one third) of stars in such clusters are
{\it triggered} to form or otherwise affected by the presence of
nearby massive stars, then a significant fraction (say, one sixth) of
all stars will have been affected by this process.  Given that our own
Solar System probably formed in a massive cluster (Hester et al.\
2004; Hester \& Desch 2005), understanding star formation in
high-mass-star-forming regions is needed to put our own formation in
context.  Accurate predictions of the initial mass function (IMF), as
well as a quantification of the timing and conditions of
protoplanetary disk growth, rely on identifying processes by which
massive stars can affect subsequent star formation and quantifying the
extent to which this has occurred in high-mass-star-forming regions.
Is star formation in \hii~region environments coeval?  If it is not,
is later star formation physically connected to the presence of
high-mass stars?  If so, by what mechanism?

A number of mechanisms by which massive stars can affect the
subsequent star formation in a region have been proposed.  Here we
review three.  One proposed mechanism for triggered star formation is
the ``radiatively-driven implosion" (RDI) model, which invokes
photoionizing ultraviolet (UV) radiation from massive stars feeding
back into the environment of the nearby molecular cloud and
interstellar medium (ISM).  In this model, pre-existing molecular
clumps in the cloud become surrounded on all sides by high-pressure
ionized gas heated by the photoionizing UV.  The pressure that is
exerted on the surface of a molecular clump leads to the implosion of
the clump and to the formation of a cometary globule (Bertoldi 1989;
Bertoldi $\&$ McKee 1990; Larosa 1983).  The collapse phase for this
scenario is relatively short (i.e., less than the typical free-fall
timescale $t_{\rm ff} \, \sim \, 3 \times 10^{5} \, {\rm yr}$), and
during this phase a shock front (traveling behind the ionization
front) will progress into the clump, leading to the formation of a
dense core within it (Deharveng et al.\ 2005).  One observational
signature of this mode of triggered star formation would be cometary
globules having dense heads and a tail extending away from the
ionizing source.  A key prediction of this model is that there should
be no newly forming stars located ``ahead'' of the ionization front,
in gas that has not yet been radiatively heated.

A second proposed mechanism is the ``collect and collapse'' (C\&C)
model first proposed by Elmegreen \& Lada (1977).  This model invokes
the standard picture of a slow-moving D-type ionization front with
associated shock front that proceeds the ionization front (see
Osterbrock 1989).  Dense gas tends to pile up in the space between the
shock and ionization fronts, with the surface density of this layer
increasing over time.  The ``collect and collapse" model hypothesizes
that at some critical threshold (as the layer gets denser and as it
cools) the dense layer becomes gravitationally unstable and fragments
into dense clumps that then go on to form stars.  The attraction of
this model is that both large numbers of stars can form in this layer,
as well as massive stars (Whitworth et al.\ 1994).  Because
fragmentation in the dense layer, driven by gravitational instability,
grows fastest for long wavelengths, an observational test of the C\&C
scenario is that large fragments typically will be evenly spaced (and
detectable in mid-infrared [mid-IR] and millimeter emission) within a
shell that surrounds the \hii~region (Deharveng et al.\ 2005).

A third proposed mechanism, proposed by Hester et al.\ (1996) in the
context of M16, also invokes a slow-moving D-type ionization front
with associated shock, like the C\&C scenario.  In this scenario,
clumps are triggered to form and collapse by the high pressures in the
shocked gas ahead of the ionization front.  Dense clumps were probably
not pre-existing in the gas, but were either formed from marginally
stable overdense regions in the molecular cloud, or created by a
fragmentation instability as in the C\&C model.  As in the RDI model,
the high pressures of the compressed gas lead to rapid (i.e., $<
t_{\rm ff} \approx 3 \times 10^5 \, {\rm yr}$) collapse of the clumps
(Garc\'{i}a-Segura $\&$ Franco 1996).  As the ionization front
advances through the region, the gas surrounding clumps is exposed,
leading to objects like cometary globules, which Hester et al.\ (1996)
termed ``evaporating gaseous globules", or EGGs.  Later, as more gas
is photoevaporated, these evaporated clumps are observed as
``proplyds" or protoplanetary disk.  Because the clumps are triggered
to collapse by the high pressures in the post-shock region, a key
observational test of this scenario is that Young Stellar Objects
(YSOs) will be spatially correlated with ionization fronts within
\hii~region; unlike in the C\&C scenario, though, protostars will not
necessarily be regularly spaced.  Another key test of this scenario is
that the evolutionary stage (age) of clumps and subsequent YSOs and
disks should increase with distance from the shock front.

While each of the scenarios makes distinct observational tests, it is
difficult to constrain which scenario is operating.  For example,
proplyds (as predicted by the third scenario) are observed in a
variety of regions, including M16 (Hester et al.\ 1996), W3/W4 (Oey et
al.\ 2005), RCW 49 (Whitney et al.\ 2004a), and M20 (Cernicharo et
al.\ 1998; Hester et al.\ 1999).  But the existence of proplyds is not
unique to this scenario and would arise in the C\&C model as well (and
probably the RDI, too).  A sequence of star formation has been
observed by Lee \& Chen (2007) in the Ori OB1 and Lac OB1
associations, and a recent study of massive YSOs and ongoing star
formation in the LMC by Book et al.\ (2009) shows evidence for
triggered star formation of massive YSOs; but the C\&C and RDI
scenarios are not distinguished.  Further progress must rely on finer
observational tools.

As the above discussion indicates, an important diagnostic tool for
testing predictions of triggered star formation models is a measure of
the ages of detected protostellar objects.  YSOs have infrared
excesses (above the stellar photospheric emission) due to a large
infalling envelope in the early stages of formation, and from an
accretion disk in the later stages.  Traditionally, YSOs have been
separated into the ``Class I, II, III'' evolutionary stages, first
defined by Adams, Lada, and Shu (1987), based on the slopes of their
spectral energy distributions (SEDs) in the near-IR; however,
selecting YSO sources from their \textit{Spitzer} mid-IR colors may be
a more robust method (Allen et al.\ 2004; Whitney et al.\ 2003a, 2003b
\& 2004b), mainly because not all YSOs have an excess in the near-IR.
Recent studies, including Poulton et al.\ (2008), Indebetouw et al.\
(2007), and Simon et al.\ (2007), have also successfully used an
SED-fitting tool from Robitaille et al.\ (2007) to identify and
classify YSOs in multiple regions.  The online SED fitter uses a grid
of YSO models based on radiative transfer codes from Whitney et al.\
(2003a, 2003b,$\&$ 2004b).  In this paper, we use the online SED
fitting tool to confirm the candidate YSOs and to estimate their
physical properties, specially their ages.

The \hii~region NGC 2467, also known as Sharpless 311, is located at a
distance of 4.1 kpc (Feinstein $\&$ V\'{a}zquez 1989).  This region is
dominated by one O6 Vn star, HD 64315.  There are also two stellar
clusters in the area, Haffner 19 (H19) and Haffner 18ab (H18ab), that
contain one later type O and additional B stars, but most ($\sim
70\%$) of the ionizing radiation comes from the O6 star.  Using UBVRI
broad-band data, Moreno-Corral et al.\ (2002) determined that there
are 34 B stars in the cluster H19, the most massive being a B1V star.
They estimate the age of H19 to be $\approx \, 2 \, {\rm Myr}$.
Fitzgerald $\&$ Moffat (1974) estimate the age of H18ab to only be 1
Myr, and Munari et al.\ (1998) obtained a best-fit age for H18ab of 2
Myr. Pismis and Moreno (1976) claim that the entire \hii~region
complex is very young, with an age around 2 - 3 Myr.  We will assume
an age for NGC 2467 of 2 Myr.  Recently, De Marco et al.\ (2006) used
\textit{Hubble Space Telescope} (\textit{HST}) Advanced Camera for
Surveys (ACS) data to identify a large number of brightened ridges and
cloud fragments in NGC 2467.  The ionization front appears to be very
near to many of these cloud fragments, suggesting that they have
recently broken off from the molecular cloud as they are uncovered by
the advancing ionization front.  In the process these fragments are
being photoevaporated, and thus NGC 2467 is an excellent candidate for
a study of star formation in an \hii~region environment, and to test
the predictions of the triggered star formation scenarios outlined
above.

In this paper, we use \textit{Spitzer} observations to locate
protostars in NGC 2467.  We then use their SEDs to confirm the
candidates as protostellar objects, and to determine their physical
properties, especially their age.  These observations allow us to
constrain the timescales for their formation and the processes at play
during their formation.  We also quantify the overall distribution of
these objects and their spatial correlation with tracers of star
formation with and massive OB stars in the region.  We then test
whether the formation of stars in this \hii~region was coeval or at
some level triggered, and attempt to distinguish between the various
mechanisms proposed for triggered star formation.

\section{Observations and Data Reduction}

The data for NGC 2467 were obtained by the \textit{Spitzer Space
Telescope} with the Infrared Array Camera (IRAC; Fazio et al.\ 2004)
and by the Multiband Imaging Photometer for \textit{Spitzer} (MIPS;
Rieke et al.\ 2004) during Cycle $\#$2 as part of the \textit{Spitzer}
program PID 20726.  We have data in all four IRAC wavelength bands and
in the 24 $\mu$m band with MIPS.  The four IRAC bands are centered at
3.6, 4.5, 5.8, and 8.0 $\mu$m.  The total mosaicked image size for NGC
2467 is $\sim$31\farcm7 $\times$ 16\farcm3 for IRAC and
$\sim$20\farcm8 $\times$ 21\farcm3 for MIPS. The pixel scale is
1\farcs20 per pixel in all four IRAC bands and 2\farcs45 per pixel in
the MIPS 24 $\mu$m band.  The IRAC channel 1 and channel 3 bands
coincide on the sky, as do the channel 2 and channel 4 bands, but
there is an offset of 6\farcm73 between the center positions of the
two pairs. The IRAC data for each frame were exposed for 12 seconds,
and the field was dithered five times resulting in a total exposure
time of 60 seconds per pixel.  For the 24 $\mu$m data, each frame was
10 seconds long with 4 cycles, resulting in a total exposure time of
560 seconds.  The MIPS data were calibrated by the \textit{Spitzer}
Science Center (SSC) pipeline version S13.2.0, and the IRAC data were
calibrated with the SSC pipeline version S14.0.0.

Mosaicked images were re-made by using the Basic Calibrated Data
(BCDs) from the SSC pipeline in the MOPEX software program, version
030106. Within MOPEX, cosmic rays were rejected from the BCD images
and detector artifacts were removed. Point sources were extracted from
the mosaicked images by first identifying a small region with little
background nebulosity present in both the 5.8 and 8.0 $\mu$m images.
The 5.8 and 8.0 $\mu$m images have a much larger extent of nebulosity
and therefore point sources are harder to identify above the emission
compared with corresponding point sources in the other two IRAC
frames. Point sources were selected in this low-background region in
the 5.8 and 8.0 $\mu$m images by eye, and fluxes were measured for
these specific sources using the \textit{aper.pro} task in the IDLPHOT
package in IDL.  Using this same ``blank'' region, an artificial
higher background, based on the highest background level of the rest
of the field, was added to the small ``blank'' region and again
sources were identified by eye.  This was done in order to help
constrain the minimum flux level needed to detect sources in the
bright background regions of the dataset.  Using only the sources
still detected over the artificially applied background level, we
determined a minimum flux level of these sources and set this as the
minimum flux threshold for reliable detection of point sources
throughout the entire region, regardless of the actual background
level. The calculated minimum flux levels are 0.5, 0.4, 0.8, and 1.0
mJy for IRAC channels 1, 2, 3, and 4, respectively. This corresponds
to minimum stellar masses of 0.2, 0.2, 0.3, and 0.3 M\sol~for each of
the four IRAC bands, based on the fluxes from the YSO SED models of
Robitaille et al.\ (2006).   

The task \textit{find.pro}, in IDLPHOT, was run to extract point
sources from all of the mosaicked images by selecting sources using
the minimum flux threshold for reliable identification, described
above. While this excluded some fainter sources in outlying
low-brightness regions (sources that may have been identifiable, but
are below the minimum flux threshold), the results more accurately
represent the actual source distribution in the region of NGC 2467,
helping to exclude faint background and foreground stars.  There are
also more detectable point sources in the IRAC ch1 and ch2 bands than
there are in the longer wavelength IRAC bands (ch3 and ch4). However,
for this analysis we are interested in selecting YSOs based on their
mid-IR excess in \textit{Spitzer} IRAC colors using all four bands,
therefore we only select sources that can be detected in all IRAC
channels.

Aperture photometry was performed on
the extracted sources using \textit{aper.pro} in IDLPHOT.  For the
IRAC images, apertures of radii 3 pixels (3\farcs6) were used with a
background sky annulus of 10-20 pixels (12-24\arcs).  For the 24
$\mu$m images, an aperture of 2.5 pixels (6\arcs) was used with a sky
annulus of 2.5-5.5 pixels (6-13\arcs).  Aperture corrections of
1.112, 1.113, 1.125, 1.218, and 1.698 and zero-magnitude fluxes of
280.9 Jy, 179.7 Jy, 115.0 Jy, 64.13 Jy, and 7.14 Jy, provided by the
SSC, were applied for IRAC channels 1, 2, 3, 4, and MIPS 24 $\mu$m
respectively.  Magnitude errors were calculated for each source in
each band using the standard method described by Everett $\&$ Howell
(2001).  Sky fluxes were determined by the aperture photometry routine, and
the gain and readnoise of each passband were provided by the SSC.
Magnitudes and the associated errors of our sources are reported in Table 1.

\section{Results} 
We found 186 sources in NGC 2467 that were detected
above the minimum flux values in each of the four IRAC bands, and we
found 23 MIPS 24 $\mu$m point sources. Color-color plots were
generated from IRAC and MIPS fluxes using the [3.6]-[4.5],
[5.8]-[8.0], [3.6]-[5.8] and the [8.0] - [24.0] colors. From these
plots, we identified more than 50 sources with infrared excesses in
one or more mid-IR colors.  The infrared excesses of these sources
indicate that they are a young population.  The color criteria for the
different protostellar objects are listed in Table 2.  The IRAC color
criteria are taken from Megeath et al.\ (2004) and are based on models
from Allen et al.\ (2004) and Whitney et al.\ (2003a $\&$ 2004b) of
the youngest protostellar objects with accreting envelopes (Class I/0)
and young low-mass stars with disks (Class II). When 
%\begin{landscape}

\begin{deluxetable*}{cccccrrrrr} [h!]
\tablecolumns{10}
\tablewidth{0pc}
\tablecaption{\textit{Spitzer} Photometry for NGC 2467}
\tablehead{ 
\colhead {Source} & \colhead{IRAC} & \colhead{MIPS} & \colhead{R.A.} & \colhead{Decl.} & \colhead{[3.6]} & \colhead{[4.5]} &
\colhead{[5.8]} & \colhead{[8.0]} & \colhead{[24]}\\ \colhead{$\#$}& \colhead{Class} & \colhead{Class} & \colhead{J2000} & \colhead{J2000} & \colhead{} & \colhead{} & \colhead{} & \colhead{} & \colhead{}} 
\startdata 
1&I/0 & \nodata & 7 52 34.45 & -26 26 34.65 & 11.31 $\pm$ 0.03 & 10.84 $\pm$ 0.03 &  9.82 $\pm$ 0.09 & 8.38 $\pm$ 0.04 & \nodata \\ 
2&I/0 & \nodata & 7 52 34.43 & -26 26 43.83 & 12.07 $\pm$ 0.05 & 11.52 $\pm$ 0.05 & 10.03 $\pm$ 0.10 & 8.55 $\pm$ 0.05 & \nodata  \\
3&I/0 & I/0 & 7 52 36.77 & -26 23 23.51 & 11.35 $\pm$ 0.03 & 10.76 $\pm$ 0.02 & 9.31 $\pm$ 0.05 & 7.80 $\pm$ 0.02 & 5.31 $\pm$ 0.18\\
4&I/0 & \nodata & 7 52 38.00 & -26 21 31.91 & 12.35 $\pm$ 0.05 & 11.65 $\pm$ 0.04 & 10.54 $\pm$ 0.12 & 9.22 $\pm$ 0.06 & \nodata \\
5&I/0 & I/0 & 7 52 36.44 & -26 26 11.35 & 11.09 $\pm$ 0.02 & 10.52 $\pm$ 0.02 &  9.38 $\pm$ 0.06 & 8.07 $\pm$ 0.03 & 4.97 $\pm$ 0.12 \\
6&I/0 & \nodata & 7 52 39.99 & -26 25 32.47 & 12.51 $\pm$ 0.06 & 11.97 $\pm$ 0.06 & 10.70 $\pm$ 0.12 & 9.12 $\pm$ 0.05 & \nodata \\
7&I/0 & \nodata & 7 52 44.41 & -26 22 59.32 & 13.58 $\pm$ 0.17 & 12.60 $\pm$ 0.10 & 12.07 $\pm$ 0.52 &11.11 $\pm$ 0.39 & \nodata  \\
8&I/0 & \nodata & 7 52 45.29 & -26 24 22.22 & 11.72 $\pm$ 0.05 & 11.17 $\pm$ 0.04 & 10.10 $\pm$ 0.14 & 8.79 $\pm$ 0.08 & \nodata  \\
9&I/0 & I/0 & 7 52 45.00 & -26 24 27.83 & 12.61 $\pm$ 0.10 & 10.44 $\pm$ 0.02 &  9.17 $\pm$ 0.06 & 8.38 $\pm$ 0.05 & 3.15 $\pm$ 0.02  \\
10&I/0 & I/0 & 7 52 46.68 & -26 23 59.79 & 12.62 $\pm$ 0.09 & 10.96 $\pm$ 0.03 &  9.81 $\pm$ 0.08 & 8.65 $\pm$ 0.05 & 4.94 $\pm$ 0.07 \\
11&I/0 & \nodata & 7 52 47.87 & -26 22 23.17 & 12.75 $\pm$ 0.08 & 12.23 $\pm$ 0.07 & 12.31 $\pm$ 0.64 &10.48 $\pm$ 0.21 & \nodata \\
12&I/0 & \nodata & 7 52 52.82 & -26 15 17.20 & 14.23 $\pm$ 0.29 & 13.10 $\pm$ 0.15 & 11.87 $\pm$ 0.44 &10.17 $\pm$ 0.15 & \nodata  \\
13&I/0 & II & 7 52 51.65 & -26 25 42.72 & 11.17 $\pm$ 0.02 & 10.49 $\pm$ 0.02 &  9.82 $\pm$ 0.05 & 7.96 $\pm$ 0.02 & 4.85 $\pm$ 0.03 \\
14&I/0 & \nodata & 7 53  1.60 & -26 19 59.20 & 12.33 $\pm$ 0.04 & 11.84 $\pm$ 0.04 & 11.36 $\pm$ 0.20 &10.24 $\pm$ 0.10 & \nodata  \\
15&I/II & II & 7 52 16.33 & -26 23 14.46 & 11.14 $\pm$ 0.02 & 11.06 $\pm$ 0.03 & 10.58 $\pm$ 0.14 & 9.05 $\pm$ 0.06 & 3.18 $\pm$ 0.02  \\
16&I/II & I/0 & 7 52 20.15 & -26 27 54.10 & 11.21 $\pm$ 0.03 & 11.20 $\pm$ 0.05 &  8.53 $\pm$ 0.03 & 6.78 $\pm$ 0.01 & 3.05 $\pm$ 0.04  \\
17&I/II & \nodata & 7 52 21.33 & -26 27 10.30 & 11.63 $\pm$ 0.04 & 11.62 $\pm$ 0.06 & 10.92 $\pm$ 0.23 & 9.74 $\pm$ 0.14 & \nodata \\
18&I/II & \nodata & 7 52 24.04 & -26 25 28.96 & 10.63 $\pm$ 0.02 & 10.61 $\pm$ 0.02 &  9.96 $\pm$ 0.08 & 8.78 $\pm$ 0.05 & \nodata  \\
19&I/II & I/0 & 7 52 27.53 & -26 27 21.17 & 12.73 $\pm$ 0.09 & 12.56 $\pm$ 0.13 & 10.60 $\pm$ 0.14 & 8.81 $\pm$ 0.05 & 5.04 $\pm$ 0.09 \\
20&I/II & \nodata & 7 52 31.64 & -26 22 07.19 & 12.07 $\pm$ 0.04 & 11.90 $\pm$ 0.06 & 11.18 $\pm$ 0.24 & 9.99 $\pm$ 0.14 & \nodata \\
21&I/II & \nodata & 7 52 35.54 & -26 25 08.30 & 11.92 $\pm$ 0.04 & 11.71 $\pm$ 0.06 & 10.19 $\pm$ 0.10 & 8.67 $\pm$ 0.04 & \nodata \\
22&I/II & I/0 & 7 52 36.38 & -26 25 56.30 & 11.13 $\pm$ 0.02 & 10.83 $\pm$ 0.02 &  9.10 $\pm$ 0.03 & 7.47 $\pm$ 0.01 & 4.70 $\pm$ 0.10 \\
23&I/II & II & 7 52 38.58 & -26 23 09.94 & 11.47 $\pm$ 0.03 & 11.37 $\pm$ 0.04 & 10.59 $\pm$ 0.13 & 9.32 $\pm$ 0.07 & 5.39 $\pm$ 0.11 \\
24&I/II & \nodata & 7 52 43.16 & -26 17 15.79 & 12.76 $\pm$ 0.07 & 12.50 $\pm$ 0.08 & 11.85 $\pm$ 0.40 &10.33 $\pm$ 0.17 & \nodata \\
25&I/II & I/0 & 7 52 40.95 & -26 23 48.73 & 11.28 $\pm$ 0.02 & 11.20 $\pm$ 0.03 &  9.80 $\pm$ 0.06 & 8.31 $\pm$ 0.03 & 5.31 $\pm$ 0.10 \\
26&I/II & II & 7 52 44.53 & -26 17 26.95 & 11.68 $\pm$ 0.03 & 11.42 $\pm$ 0.03 & 10.30 $\pm$ 0.10 & 8.65 $\pm$ 0.04 & 4.31 $\pm$ 0.04 \\
27&I/II & \nodata & 7 52 42.22 & -26 22 51.03 & 12.55 $\pm$ 0.07 & 12.16 $\pm$ 0.07 & 11.51 $\pm$ 0.29 &10.32 $\pm$ 0.17 & \nodata \\
28&I/II & I/0 & 7 52 42.88 & -26 24 12.95 & 10.98 $\pm$ 0.03 & 10.80 $\pm$ 0.03 &  9.39 $\pm$ 0.06 & 7.71 $\pm$ 0.02 & 2.19 $\pm$ 0.06 \\
29&I/II & II & 7 52 42.82 & -26 25 45.80 & 11.09 $\pm$ 0.02 & 10.72 $\pm$ 0.02 & 10.21 $\pm$ 0.08 & 8.88 $\pm$ 0.04 & 5.05 $\pm$ 0.03 \\
30&I/II & \nodata & 7 52 49.75 & -26 16 33.06 & 11.11 $\pm$ 0.02 & 11.03 $\pm$ 0.03 & 10.39 $\pm$ 0.11 & 9.11 $\pm$ 0.05 & \nodata \\
31&I/II & I/0 & 7 52 50.22 & -26 18 07.62 & 12.61 $\pm$ 0.07 & 12.22 $\pm$ 0.08 & 10.12 $\pm$ 0.10 & 8.42 $\pm$ 0.04 & 3.75 $\pm$  0.02 \\ 
32&I/II & II & 7 52 50.25 & -26 26 40.30 & 11.39 $\pm$ 0.02 & 11.35 $\pm$ 0.03 & 10.95 $\pm$ 0.14 & 9.27 $\pm$ 0.05 & 5.42 $\pm$ 0.04 \\
33&II & DD & 7 52 17.82 & -26 25 20.77 & 11.17 $\pm$ 0.03 & 11.14 $\pm$ 0.04 & 11.40 $\pm$ 0.32 & 10.81 $\pm$ 0.35 & 2.68 $\pm$ 0.06 \\
34&II & \nodata & 7 52 23.29 & -26 26 50.30 & 10.24 $\pm$ 0.01 & 10.22 $\pm$ 0.02 & 10.05 $\pm$ 0.09 &  9.33 $\pm$ 0.08 & \nodata \\
35&II & \nodata & 7 52 32.01 & -26 22 37.26 & 12.49 $\pm$ 0.06 & 12.15 $\pm$ 0.07 & 11.58 $\pm$ 0.33 & 10.70 $\pm$ 0.26 & \nodata \\
36&II & \nodata & 7 52 32.14 & -26 22 30.21 & 11.99 $\pm$ 0.04 & 11.58 $\pm$ 0.04 & 11.20 $\pm$ 0.24 & 10.52 $\pm$ 0.23 & \nodata \\
37&II & \nodata & 7 52 35.86 & -26 21 58.04 & 10.79 $\pm$ 0.02 & 10.28 $\pm$ 0.02 &  9.94 $\pm$ 0.08 &  9.44 $\pm$ 0.09 & \nodata \\
38&II & II & 7 52 33.51 & -26 26 47.88 &  9.05 $\pm$ 0.01 &  8.43 $\pm$ 0.01 &  7.67 $\pm$ 0.01 &  6.65 $\pm$ 0.01 & 3.15 $\pm$ 0.02 \\
39&II & DD & 7 52 35.54 & -26 25 44.39 & 11.92 $\pm$ 0.04 & 11.85 $\pm$ 0.06 & 11.90 $\pm$ 0.44 & 10.92 $\pm$ 0.32 & 4.45 $\pm$ 0.04 \\
40&II & II & 7 52 35.54 & -26 26 34.36 &  9.21 $\pm$ 0.01 &  8.56 $\pm$ 0.01 &  7.92 $\pm$ 0.02 &  7.03 $\pm$ 0.01 & 4.74 $\pm$ 0.11 \\
41&II & \nodata & 7 52 36.23 & -26 25 48.15 & 12.02 $\pm$ 0.04 & 11.53 $\pm$ 0.04 & 11.19 $\pm$ 0.23 & 10.54 $\pm$ 0.22 & \nodata \\
42&II & \nodata & 7 52 39.13 & -26 23 21.85 & 10.63 $\pm$ 0.01 & 10.58 $\pm$ 0.02 & 10.19 $\pm$ 0.09 &  9.19 $\pm$ 0.06 & \nodata \\
43&II & \nodata & 7 52 47.20 & -26 15 58.44 & 10.59 $\pm$ 0.01 & 10.19 $\pm$ 0.02 &  9.63 $\pm$ 0.06 &  8.98 $\pm$ 0.05 & \nodata  \\
44&II & \nodata & 7 52 43.79 & -26 24 21.94 & 10.94 $\pm$ 0.03 & 10.26 $\pm$ 0.02 &  9.53 $\pm$ 0.08 &  8.47 $\pm$ 0.06 & \nodata \\
45&II & \nodata & 7 52 55.88 & -26 22 39.50 & 11.21 $\pm$ 0.02 & 10.83 $\pm$ 0.02 & 10.54 $\pm$ 0.12 & 10.00 $\pm$ 0.12 & \nodata \\

\enddata 
\end{deluxetable*} 
%\end{landscape}

%\begin{landscape}

\begin{deluxetable*}{ccccccc}
\tablecolumns{7}
\tablewidth{0pc}
\tablecaption{Color Criteria for Selection of Protostellar Objects}
\tablehead{
\colhead {IRAC} & \colhead{IRAC} & \colhead{IRAC} & \colhead{MIPS} & \colhead{MIPS}& \colhead{AGB}  & \colhead{Extra-Gal}\\\colhead{Class I/0} & \colhead{Class I/II} & \colhead{Class II} & \colhead{Class I/0} & \colhead{Class II} & \colhead{} & \colhead{}}
\startdata
[3.6]-[4.5] $\geq$ 0.8 & [3.6]-[4.5] $\leq$ 0.4 & [3.6]-[4.5] $\leq$ 0.8 &  [3.6]-[5.8] $\geq$ 1.4  & [3.6]-[5.8] $\leq$ 1.4 & [8.0]: 3 - 9 & [8.0] $\leq$ 14 - ([4.5]-[8.0])\\
or & and & and & and & and & and & \nodata\\
$[$5.8]-[8.0] $\geq$ 1.1 & [5.8]-[8.0] $\geq$ 1.1 & 0.4 $\leq$ [5.8]-[8.0] $\leq$ 1.1 & [8.0]-[24]: 2.0 - 6.0 & [8.0]-[24]: 2.0 - 6.0 & [4.5]-[8.0] $\leq$ 1 & \nodata\\
$\&$ [3.6]-[4.5] $\geq$ 0.4 & & & & & &\\
\enddata 
\end{deluxetable*} 
%\end{landscape}

\begin{deluxetable}{cccccc}
\tablecolumns{6}
\tablewidth{0pc}
\tablecaption{\textit{Spitzer} IRAC Colors and 2MASS Photometry for NGC 2467}
\tablehead{ 
\colhead {Source $\#$} & \colhead{[3.6]-[4.5]} & \colhead{[5.8]-[8.0]} & \colhead{J} & \colhead{H} & \colhead{K}} 
\startdata 
1 & 0.47& 1.44& 14.58 & 13.42 & 12.61 \\
2 & 0.56& 1.47& 16.54 & 15.34 & 14.17 \\
3 & 0.60 & 1.51& 16.54 & 14.66 & 13.42 \\
4 & 0.70& 1.32& 15.74 & 14.65 & 13.87 \\
5 & 0.57 & 1.31& 15.31 & 13.88 & 12.85 \\
6 & 0.54& 1.59& 14.66 & 13.89 & 13.32 \\
7 & 0.98& 0.96& \nodata & \nodata & \nodata \\
8 & 0.55& 1.32& 16.52 & 14.56 & 13.47 \\
9 & 2.17& 0.79& \nodata & \nodata & \nodata \\
10& 1.66& 1.16& \nodata & \nodata & \nodata \\
11& 0.52& 1.83& 15.15 & 14.15 & 13.60 \\
12& 1.13& 1.70& \nodata & \nodata & \nodata\\
13& 0.67& 1.85& 14.56 & 13.81 & 12.85 \\
14& 0.49& 1.12& 14.98 & 14.14 & 13.66 \\
15& 0.08& 1.53& 11.37 & 11.28 & 11.15 \\
16& 0.01& 1.75& 14.66 & 14.52 & 14.48 \\
17& 0.00 & 1.18& 12.12 & 11.82 & 11.77 \\
18& 0.02 & 1.18& 11.13 & 10.79 & 10.73 \\
19& 0.16& 1.79& 15.21 & 14.81 & 14.42 \\
20& 0.16& 1.19& 13.53 & 12.98 & 12.57 \\
21& 0.21& 1.53& 14.75 & 13.71 & 13.16 \\
22& 0.30& 1.63& 15.67 & 14.24 & 12.53 \\
23& 0.09& 1.28& 12.19 & 11.89 & 11.68 \\
24& 0.26& 1.51& 14.89 & 14.30 & 13.83 \\
25& 0.08& 1.49& 12.54 & 12.09 & 11.83 \\
26& 0.26& 1.65& 12.00 & 11.94 & 11.98 \\
27& 0.39& 1.18& 15.20 & 13.79 & 13.23 \\
28& 0.19& 1.68& 11.46 & 11.36 & 11.31 \\
29& 0.37& 1.32& 14.44 & 13.64 & 12.79 \\
30& 0.08& 1.27& 11.85 & 11.33 & 11.14 \\
31& 0.39& 1.70& \nodata & \nodata & \nodata \\ 
32& 0.04& 1.68& 11.77 & 11.60 & 11.43 \\
33& 0.03& 0.59& 11.88 & 11.34 & 11.19 \\
34& 0.02& 0.72& 11.01 & 10.38 & 10.26 \\
35& 0.34& 0.88& 15.37 & 14.17 & 13.60 \\
36& 0.41& 0.68& 15.08 & 13.99 & 13.24 \\
37& 0.51& 0.49& 13.12 & 12.26 & 11.67 \\
38& 0.62& 1.02& 12.93 & 11.62 & 10.53 \\
39& 0.08& 0.98& 12.32 & 12.12 & 12.02 \\
40& 0.66& 0.89& 13.09 & 11.77 & 10.63 \\
41& 0.49& 0.64& 12.12 & 11.82 & 11.77 \\
42& 0.05& 1.00& 10.99 & 10.85 & 10.72 \\
43& 0.40& 0.65& 13.22 & 12.76 & 12.15 \\
44& 0.68& 1.07& \nodata & \nodata & \nodata \\
45& 0.38& 0.54& 13.51 & 12.71 & 12.18 \\
\enddata 
\end{deluxetable} 
 
\noindent collectively
discussing the population of young objects in this region we refer to
them as both YSOs and protostellar objects.  Possible Class II YSOs
and Class I/0 protostars were also identified from the MIPS 24 $\mu$m
sources based on color criteria from Reach et al.\ (2004). Similar
methods were used by Rho et al.\ (2006) to identify protostars in the
Trifid Nebula with \textit{Spitzer}.  In total, 46 possible
protostellar candidates were identified in NGC 2467.

The Two-Micron All Sky Survey (2MASS) All-Sky Point Source Catalog
(PSC) was used to find 2MASS counterparts to the \textit{Spitzer}
sources. There are 166 point sources that were detected in both
surveys. There are 29 2MASS sources that show an infrared excess in
2MASS colors and 27 of them correspond to an IRAC YSO source.  The
overlap of 27 of the 2MASS infrared excess sources with the
\textit{Spitzer} detected YSOs validates the methods for selecting
YSOs.

Background asymptotic giant branch (AGB) stars and extragalactic
contaminants could be present in the data since they can have similar
colors as YSOs.  The 46 protostellar candidates were first checked by
eye in all \textit{Spitzer} images in an initial attempt to rule out
any extended objects (galaxies) that might be present in the
sample. None of the 46 protostellar candidates appeared to be an
elongated or extended object. Color and magnitude criteria for both
AGB stars and extragalactic contaminants from Harvey et al.\ (2006)
were also used to attempt to reject these objects from the sample. The
\textit{Spitzer} color and magnitude criteria for AGB stars and
extragalactic sources are also listed in Table 2.  A few objects met
the AGB criteria, but most were off-cloud sources.  None of the
selected 46 protostellar objects had a magnitude or color meeting the
criteria for AGB stars. One of the selected protostellar objects
matched the criterion of an extragalactic object. This source was not
detected in the 2MASS bands, and was therefore discarded as a possible
YSO. This left a total of 45 protostellar candidates. Table 1 gives
the IRAC and/or MIPS YSO classification, position, calculated
\textit{Spitzer} magnitudes and errors for the YSOs; Table 3 lists the
IRAC colors that were used to classify the YSOs and the 2MASS
magnitudes for the 45 possible YSOs found in NGC 2467.

Figure 1 shows a three-color image of NGC 2467; the IRAC 4.5 $\mu$m
channel is in blue, the IRAC 8.0 $\mu$m channel is in green, and the
MIPS 24 $\mu$m channel is in red.  The IRAC and MIPS images show a
region of ionized gas pushing out into the surrounding molecular
cloud.  Strong polycyclic aromatic hydrocarbon (PAH) emission can be
seen in the 8.0 $\mu$m band. PAH emission is present throughout the
photodissociation region (PDR) at 8.0 $\mu$m. PAH emission shows the
locations of edges and ionization fronts created by the O star.   The
MIPS 24 $\mu$m emission is concentrated in the area surrounding the
central O6 star, indicating the presence of warm dust.  The locations
of the possible detected YSOs are shown in Figure 1, along with the
locations of known OB stars in the region.

Figure 2 shows an IRAC color-color diagram of [3.6]-[4.5]
vs. [5.8]-[8.0] for NGC 2467.  The protostellar candidates are plotted
in color in Figure 2; Class II sources are in green, Class I/II
sources are in yellow, and Class I/0 sources are plotted in red.
Figure 3 is a color-color diagram for the 23 detected point sources in
the MIPS 24 $\mu$m band, with [3.6]-[5.8] vs. [8.0]-[24].  Of these,
18 MIPS sources were classified as protostars based on their
colors. Class I/0 sources are plotted in red and Class II sources are
shown in green. Of the 18 MIPS 24 $\mu$m sources that were found to
have a mid-IR excess, all corresponded to an IRAC point source that
also had a measured color excess in one or more IRAC colors.  Sources
that had a large color excess in the [8.0]-[24] color, but a
[3.6]-[5.8] color less than zero, are defined by Reach et al.\ (2004)
as sources with possible debris disks.  Four 24 $\mu$m sources fell
into this color regime.   There are 45 sources from both the IRAC and
MIPS color-color diagrams that had an infrared excess in at least two
colors, and 18 sources had a measured color excess in four different
colors.

Fourteen IRAC and 10 MIPS sources were found to have colors indicative
of Class I/0 objects. Models of objects with these colors from Allen
et al. (2004) and Whitney et al.\ (2003a, 2003b $\&$ 2004b) correspond to
protostellar objects with infalling dusty envelopes. These are the
youngest objects in the sample.  Of the 10 MIPS Class I/0 sources,
four corresponded to IRAC Class I/0 objects, and six corresponded to
IRAC Class I/II sources.  Thirteen IRAC and 8 MIPS sources were found
to have colors indicative of Class II protostellar objects. The colors
of these objects are characteristic of models of young low-mass stars
with disks (Allen et al.\ 2004 and Whitney et al.\ 2003a, 2003b $\&$
2004b). These objects are more evolved than the Class I/0 objects. Two
MIPS Class II sources correspond to an IRAC Class II, and five of the
MIPS Class II sources correspond with an IRAC Class I/II source. The
other MIPS Class II source corresponds to an IRAC Class I/0 source.
Eighteen other IRAC sources were classified as Class I/II objects
based on their colors. The MIPS classification of YSOs overlapped
fairly well with the IRAC classification. Cases where the sources did
not correspond are mostly due to the fact that we only had two separate
classifications of MIPS sources, but three possible classes for the
IRAC sources.  

Figure 4 shows a 2MASS $\&$ \textit{Spitzer} color-color diagram of
the 2MASS detected sources.  The 29 sources to the right of the
reddening vector are highlighted in blue, indicating an infrared
excess in their 2MASS colors. The reddening vector due to interstellar
extinction has a slope of 1.3 as defined by Tapia (1981). The
corresponding IRAC YSO candidate sources are also identified; 27 of
the 29 2MASS sources with an infrared excess correspond to YSOs
selected from \textit{Spitzer}. There is some separation of the IRAC
classified sources seen in the 2MASS color-color diagram.  A large
fraction of the IRAC Class I/0 objects are the reddest sources in the
K-[3.6] color. The IRAC Class II objects lie in the middle range of
color excess for the K-[3.6] color, but most have higher H-K colors
than the Class I/II objects.

\begin{figure*}
\epsscale{1.0}
\plotone{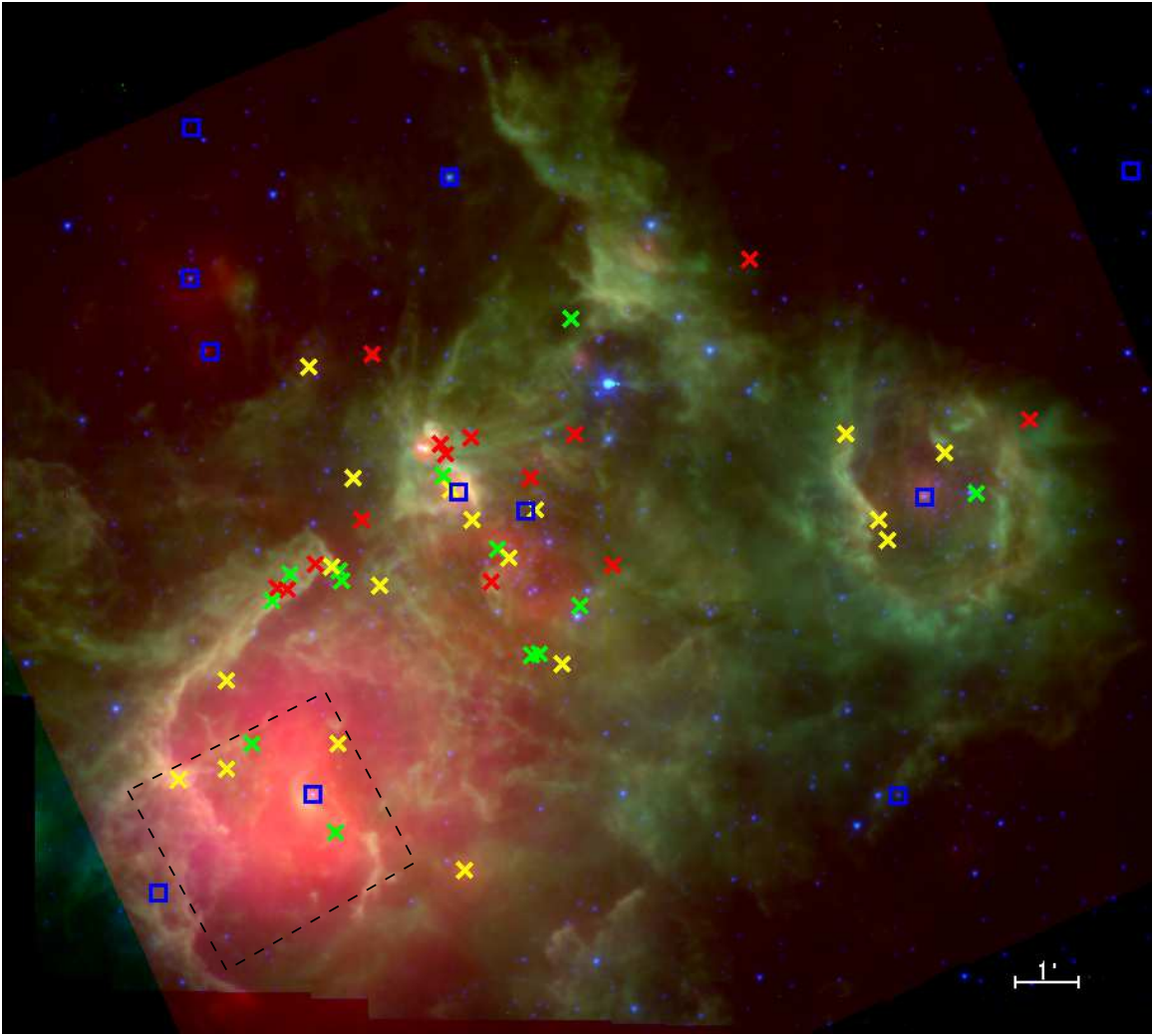}
\caption{3-color \textit{Spitzer} image of NGC 2467: Red is MIPS 24 $\mu$m, green is IRAC 8.0 $\mu$m, and blue is IRAC 4.5 $\mu$m. Red, yellow, and green Xs show locations of Class 0/I, Class I/II, and Class II protostars, respectively, and blue squares mark the locations of known OB stars in the region. The dashed line shows the location of the \textit{HST} ACS FOV, as seen in Figures 6 and 7. A scale of 1\arcm~is labeled in the figure. This line corresponds to a distance of 1.2 pc given the assumed distance of 4.1 kpc to the region.}
\end{figure*}

\begin{figure}
\epsscale{1.2}
\plotone{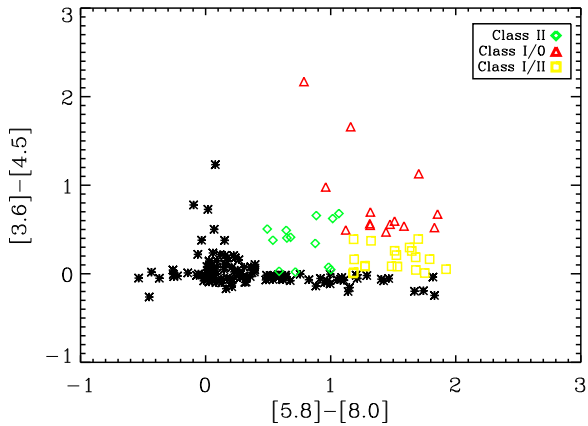}
\caption{IRAC color-color diagram for NGC 2467, showing YSO classification of IRAC detected sources. Color criteria taken from Allen et al. (2004), and Whitney et al. (2003 $\&$ 2004b).}
\end{figure}

\begin{figure}
\epsscale{1.2}
\plotone{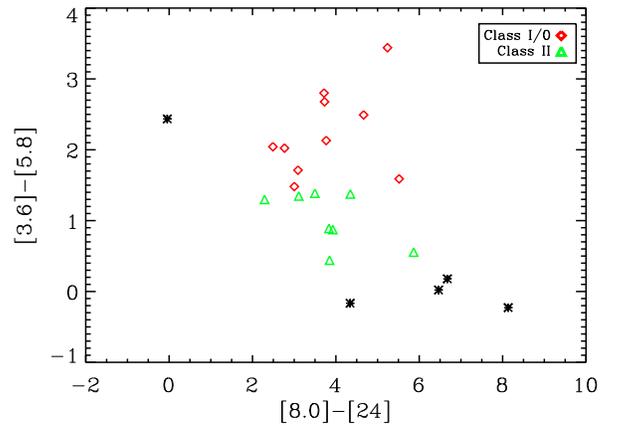}
\caption{IRAC and MIPS color-color diagram for NGC 2467, showing YSO classification of the MIPS 24 $\mu$m detected sources. Color criteria taken from Whitney et al.\ (2003a, 2003b, $\&$ 2004b) and Reach et al.\ (2004).}
\end{figure}

\section{Discussion}
\subsection{SED Model Fitting of YSOs}  

We are interested in determining the masses, ages, and other physical
properties of the detected YSOs.  To constrain the physical properties
of each of the 45 YSO candidates we used an online SED fitter from
Robitaille et al.\ (2007).  The fitter uses a grid of $2 \times
10^{4}$ SEDs corresponding to various YSO / disk models, generated by
radiative transfer codes from Whitney et al.\ (2003a, 2003b $\&$
2004b); for each YSO/disk model, SEDs are presented at 10 different
inclination angles of the disk to the line of sight, equally spaced in
intervals of cosine of the inclination, giving 2$\times$10$^{5}$ total
SEDs.  There are 14 different parameters characterizing a YSO/disk
model, including stellar mass, age, temperature, and radius, as well
as disk mass, envelope accretion rate, and other properties.  Assumed
stellar masses $M_{\star}$ range from 0.1-50 M\sol, stellar ages range
from $t_{\star} = 10^{3} - 10^{7} \, {\rm yr}$, photospheric
temperatures range from 2535 - 46000 K, and stellar radii range from
0.4 - 780 R\sol~in the models The parameters comprising the grid of
models are distributed so that there is a uniform density of models in
$\log_{10}(M_{\star})$, and close to a uniform density of models
$\log_{10}(t_{\star})$ (there is a slight bias towards larger values
of stellar age).  The reader is referred to Robitaille et al.\ (2006)
for more details on the total range of parameters.

\begin{figure}
\epsscale{1.2}
\plotone{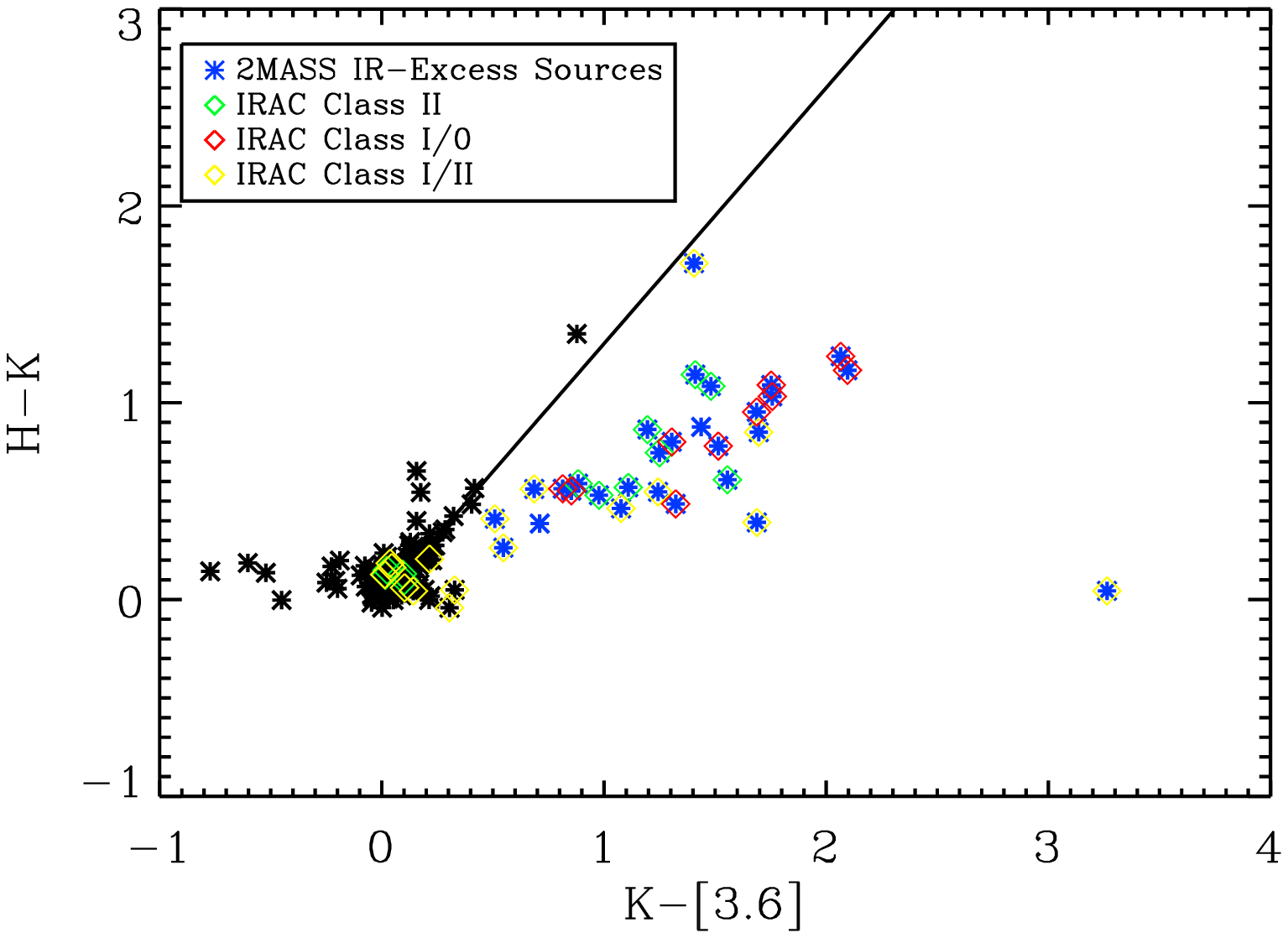}
\caption{2MASS $\&$ \textit{Spitzer} color-color diagram of the detected 2MASS sources in NGC 2467. Sources to the right of the reddening vector, which has a slope equal to 1.3, are sources with an infrared excess in the NIR. 27 out of the 29 sources with a 2MASS NIR-excess correspond to a \textit{Spitzer} selected YSO candidate.}
\end{figure}

The SED fitter from Robitaille et al.\ (2007) can help determine the
uniqueness of a given fit based on the best-fit (reduced) $\chi^{2}$ value of
a model and on the range of values for the various parameters in the
model fits. Numerous recent \textit{Spitzer} studies have successfully
used this fitter in identifying and classifying YSOs.  Simon et al.\
(2007) used the online SED fitter to identify YSOs in the \hii~region
NGC 346 in the Small Magellanic Cloud (SMC). Other recent work by
Poulton et al.\ (2008) has used the SED fitter to model detected YSOs
in the Rosette Nebula; they were able to identify over 750 YSOs and
classify their evolutionary state. Seale and Looney (2008) used the
SED fitter to look at the evolution of outflows produced by nearby
YSOs.  They found that the SED fitter from Robitaille et al.\ (2007)
adequately matches the observed data from their sample of 27 YSOs.

For NGC 2467, the \textit{Spitzer} and 2MASS fluxes for each candidate
YSO were input into the SED fitter, along with a distance estimate for
the region. All of our 45 candidate YSOs had at least four flux
measurements, with the majority (39 out of 45) having seven or eight
flux measurements. The fitter outputs a set of models for each source,
and  only models with $\chi^{2} - \chi^{2}_{\rm best} \,  \leq 3$ (per
data point) were used in our analysis in determining the best-fit
properties of each candidate YSO (similar to methods used by
Robitaille et al.\ 2007 and Simon et al.\ 2007).  We selected models
that had ages less than 2 Myr, i.e. only models with ages that are
within the assumed age of the region. We calculated the average mass
only from models that were within 3 sigma of the total $\chi$$^{2}$
from the best-fit. The 
\begin{figure}
\epsscale{1.2}
\plotone{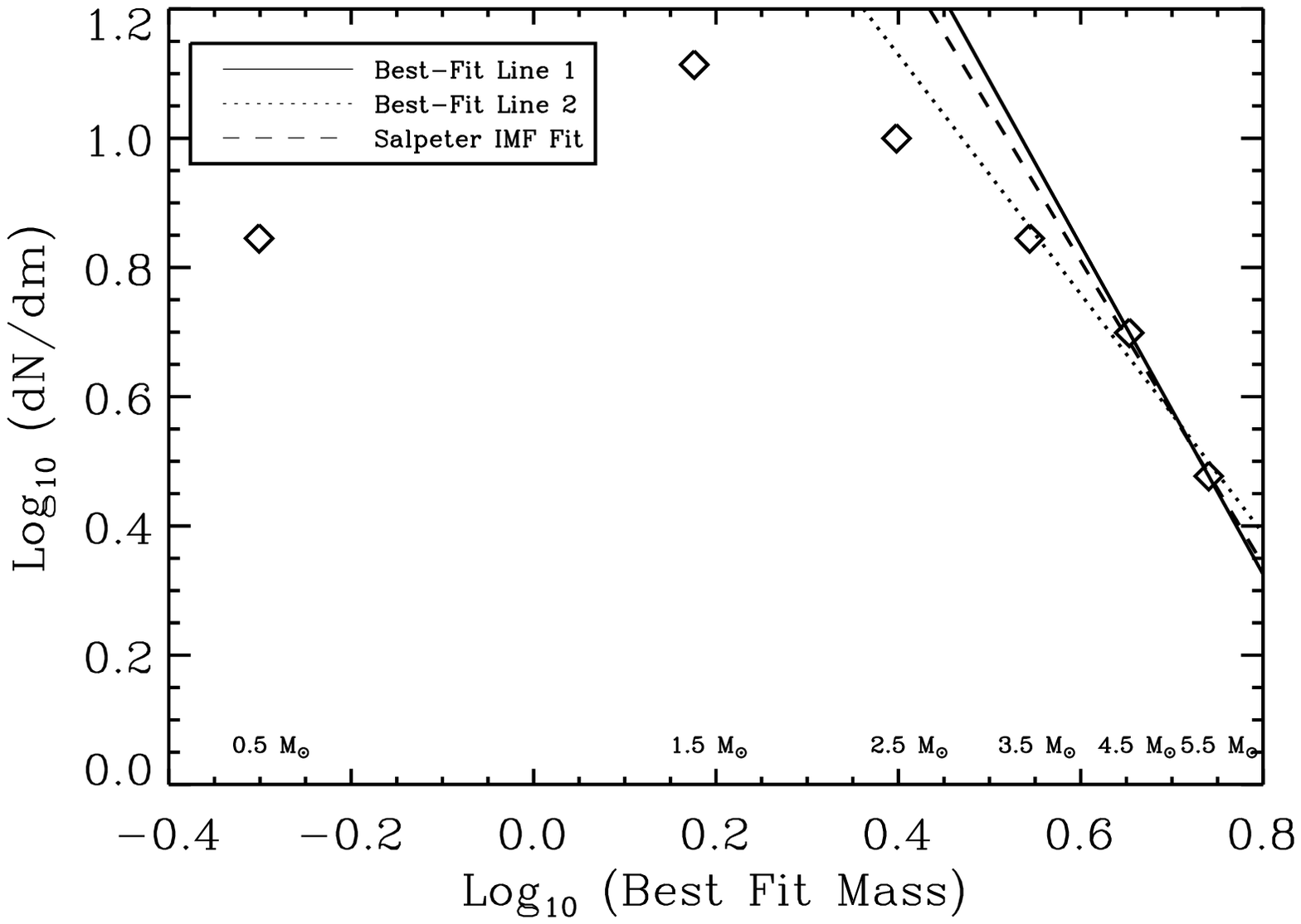}
\caption{Mass distribution function for YSOs in NGC 2467.  The solid line is the best-fit to our data through the highest two mass bins, with a slope of -2.55$\pm$0.65.  The dotted line is our second best-fit through the three highest mass bins, with a slope of -1.85$\pm$0.48.  The dashed line is for a Salpeter IMF with a slope of -2.35,  also fit to the two highest mass bins, where our sample is assumed to be more complete. The best-fit is very close to Salpeter, therefore a Salpeter IMF is assumed.}
\end{figure}
corresponding model with the lowest
$\chi$$^{2}$ and mass closest to the weighted average mass was used.
All of the candidate YSOs were checked against pure stellar photosphere
models, and the total $\chi^{2}$ of the stellar fit was compared to the
$\chi^{2}$ from the YSO fits.
None of the candidate YSOs was better fit by a pure stellar photosphere 
model, confirming our color selection criteria of YSOs for this region.  
The YSO best-fit parameters for each source are listed in Table 4.  
We found sources with masses ranging from 0.35 - 5.7 M\sol~and ages 
ranging from 10$^{3}$ - 1$\times$10$^{6}$ yr.  
There were at least seven sources that had a best-fit mass of less 
than 1 M\sol, demonstrating that we are detecting some fraction of the 
low-mass protostellar population.

\subsection{Mass Distribution of YSOs and Completeness}

A mass estimate for each of the candidate YSOs allows for an
examination of the overall population of sources in this region.  In
order to look at the distribution of masses of YSOs in NGC 2467 and to
estimate completeness, we used the best-fit model masses to determine a mass 
function for the 45 sources. We assumed a mass function of the following form:
\begin{equation}
\frac{dN}{dM}  = A (M / M_{\odot})^{-\Gamma}.
\end{equation} 
Figure 5 shows a histogram of YSOs as a function of their best-fit
mass.  The masses of the 45 YSOs were grouped into 6 equally spaced
mass bins, centered at 0.5, 1.5, 2.5, 3.5, 4.5, and 5.5 M\sol.
Various least-squares fits were performed on the data in order to
determine the exponent $\Gamma$ above.  As is apparent from Figure 5,
the deviation from a single power law becomes progressively worse at
small mass bins, indicating an incompleteness in our survey.  Our
first fit (\#1) is fit only to the two most complete mass bins (at 4.5
and 5.5 $M_{\odot}$), and yields a slope of $\Gamma = 2.55 \pm 0.65$
(where the error estimate includes only the $\sqrt{N}$ error
associated with these small-number statistics).  Our second fit (\#2)
is fit to the three highest mass bins (3.5, 4.5 and 5.5 $M_{\odot}$),
and yields a slope of $1.85 \pm 0.48$.  The first fit, using only the
highest mass bins, is definitely compatible with a Salpeter (1955)
initial mass function (IMF) with slope $\Gamma = 2.35$, which we adopt
here.

Our survey is probably close to complete in the two highest mass bins,
but probably partially incomplete in the 3.5 $M_{\odot}$ mass bin, and
increasingly incomplete in lower mass bins.  A power-law fit, it
should be noted, probably is not appropriate for the lowest mass bins
below $1 \, M_{\odot}$, which might more appropriately be modeled with
a log-normal distribution (Chabrier 2003).  This is not relevant to
the current study, however, which seeks only to judge the total number
of stars forming in the region, and not so much the exact shape of the
IMF.

\begin{deluxetable*}{cccrrrrr}
\tablecolumns{8}
\tablewidth{0pc}
\tablecaption{Best-Fit Model Parameters for YSOs in NGC 2467. YSOs models from online SED fitter by Robitaille et al.\ (2007).}
\tablehead{ 
\colhead{Source $\#$} & \colhead{Model $\#$} & \colhead{Mass} & \colhead{Age} & 
\colhead{Radius} & \colhead{Temp} & \colhead{A$_{V}$} & \colhead{Log(d)}\\
\colhead{} &\colhead{} &\colhead{(M\sol)} &\colhead{($\times$ 10$^{5}$ yr)} &\colhead{(R\sol)} &\colhead{(K)} & \colhead{(mag)} & \colhead{(kpc)}}
\startdata 
1 & 3011772 & 2.10 $\pm$ 1.17 & 1.00 & 10.98 & 4340.0 & 5.3 & 0.61 \\
2 & 3003476 & 3.59 $\pm$ 1.25 & 0.09 & 21.63 & 4298.0 & 0.6 & 0.62 \\
3 & 3004748 & 1.92 $\pm$ 0.39 & 0.07 & 14.42 & 4180.0 & 3.8 & 0.62 \\    
4 & 3018503 & 0.62 $\pm$ 0.54 & 0.14 &  6.61 & 3816.0 & 1.5 & 0.62\\
5 & 3012434 & 1.35 $\pm$ 0.59 & 0.02 &  12.91 & 4050.0 & 0.0 & 0.60 \\
6 & 3014174 & 0.99 $\pm$ 0.31 & 0.53 &  6.81 & 4103.0 & 0.8 & 0.60\\
7 & 3010612 & 0.93 $\pm$ 0.29 & 0.24 &  7.54 & 4027.0 & 20.3 & 0.61 \\    
8 & 3009627 & 2.46 $\pm$ 1.31 & 0.78 & 12.66 & 4359.0 & 4.0 & 0.60 \\
9 & 3011577 & 1.23 $\pm$ 0.44 & 0.04 & 12.89 & 3995.0 & 4.3 & 0.62 \\ 
10& 3008298 & 1.52 $\pm$ 0.75 & 0.05 & 12.33 & 4124.0 & 22.7 & 0.61 \\
11& 3016995 & 0.63 $\pm$ 0.57 & 4.70 & 3.66 & 3936.0 &  1.4 & 0.61 \\
12& 3015705 & 0.35 $\pm$ 0.46 & 0.61 &  3.72 & 3489.0 & 2.4 & 0.60 \\
13& 3006942 & 2.52 $\pm$ 1.04 & 0.22 & 14.76 & 4305.0 & 0.2 & 0.62 \\
14& 3010799 & 1.43 $\pm$ 0.90 & 0.38 &  8.76 & 4226.0 & 2.3 & 0.62 \\  
15& 3005834 & 2.37 $\pm$ 1.33 &0.56  & 12.61 & 4343.0 & 0.0 & 0.60 \\
16& 3007424 & 4.70 $\pm$ 0.33 &12.00 &  2.68 &16000.0 & 0.1 & 0.62 \\
17& 3008715 & 3.94 $\pm$ 0.93 &4.00 &  8.24 & 4903.0 & 0.0 & 0.60 \\
18& 3003892 & 4.31 $\pm$ 0.50 &6.60 &  9.46 & 5762.0 & 5.3 & 0.60 \\
19& 3016713 & 1.16 $\pm$ 0.35 &0.02  & 11.81 & 4008.0 & 0.0 & 0.60 \\
20& 3017850 & 2.50 $\pm$ 0.87 &4.40 &  5.73 & 4700.0 & 1.0 & 0.62 \\
21& 3009574 & 1.98 $\pm$ 1.57 &0.61 & 10.93 & 4313.0 & 0.0 &0.60 \\
22& 3005274 & 5.73 $\pm$ 0.26 &2.90 & 14.27 & 6109.0 & 10.4 & 0.61 \\
23& 3016205 & 3.80 $\pm$ 1.00 &3.20 &  8.62 & 4807.0 & 0.0 & 0.60 \\
24& 3010603 & 1.10 $\pm$ 0.88 &5.30 &  3.59 & 4314.0 & 1.2 & 0.60 \\
25& 3005254 & 2.55 $\pm$ 0.41 & 0.86 & 12.85 & 4369.0 & 1.8 & 0.61 \\
26& 3016360 & 1.88 $\pm$ 0.11 & 1.10& 10.02 & 4321.0 & 0.1 & 0.62 \\
27& 3002399 & 0.89 $\pm$ 0.69 &0.14&  7.86 & 3989.0 & 3.2 & 0.60 \\
28& 3017155 & 2.44 $\pm$ 1.94 &0.28  & 13.77 & 43190.0 & 0.0 & 0.60 \\
29& 3018885 & 1.92 $\pm$ 1.02 &0.92 & 10.39 & 4318.0 & 7.3 & 0.60 \\
30& 3005897 & 2.17 $\pm$ 0.85 &0.94 & 11.39 & 4341.0 & 0.0 &0.60 \\
31& 3017257 & 1.23 $\pm$ 0.27 &0.11 & 11.08 & 4057.0 & 22.4 & 0.62 \\
32& 3006533 & 4.21 $\pm$ 0.14 & 5.40&  9.21 & 5209.0 & 0.0 & 0.60 \\
33& 3011105 & 3.15 $\pm$ 0.92 & 1.30& 12.49 & 4502.0 & 1.0 & 0.62 \\
34& 3000825 & 4.17 $\pm$ 0.12 & 8.30&  8.33 & 6482.0 & 0.0 & 0.60 \\
35& 3011935 & 0.83 $\pm$ 0.33 &0.19&  7.14 & 3976.0 & 4.2 & 0.61 \\
36& 3000455 & 1.28 $\pm$ 0.94 & 1.10& 7.66 & 4214.0 & 2.0 & 0.60 \\
37& 3012794 & 4.21 $\pm$ 0.21 & 5.70&  9.26 & 5268.0 & 1.8 & 0.60 \\
38& 3005266 & 5.12 $\pm$ 0.11 & 5.70&  7.81 & 9534.0 & 5.9 & 0.61 \\
39& 3016265 & 3.81 $\pm$ 0.78 & 9.20&  7.91 & 5612.0 & 0.7 & 0.60 \\ 
40& 3005266 & 5.12 $\pm$ 0.56 & 5.70&  7.81 & 9534.0 & 6.9 & 0.62 \\
41& 3015096 & 1.40 $\pm$ 0.67 & 2.00&  7.05 & 4292.0 & 1.5 & 0.61\\
42& 3019036 & 3.62 $\pm$ 0.95 &0.59 & 17.12 & 4414.0 & 1.2 & 0.62 \\
43& 3005949 & 3.99 $\pm$ 0.45 & 10.00& 7.18 & 7158.0 & 5.5 & 0.62 \\
44& 3011228 & 2.90 $\pm$ 1.22 &0.07 & 19.35 & 4246.0 & 12.6 & 0.62 \\
45& 3010779 & 2.55 $\pm$ 0.80 &1.00 & 12.55 & 4386.0 & 2.1 & 0.62 \\
\enddata 
\end{deluxetable*}

There are approximately 20 known B8 and B9 stars in the region from
cluster surveys of H19, H18, and SH-311 (Moreno-Corral et al.\ 2002).
There are also two known O stars (O6 and O7) associated with the
region.  The number of B8/B9 stars was used to fix the IMF
normalization constant $A$ above, assuming a mass range of $3.3 - 3.8
\, M_{\odot}$; using the two known O stars (and a mass range of $30 -
40 \, M_{\odot}$ also gives us a very similar normalization constant.
The calculated average value of the normalization constant is $800 \pm
40$, again assuming Poisson statistics on the 20 B9/B8 stars.  We
therefore estimate the total number of stars between $0.2$ and $40 \,
M_{\odot}$ in NGC 2467 to be
\begin{equation}
N_{tot} = 8 \times 10^{2} \, \int_{0.2M_{\odot}}^{40M_{\odot}} (M / M_{\odot})^{-2.35} dM 
\approx 5000 \pm 1000.
\end{equation}

In order to address completeness, we estimate the total fraction of
YSOs that should have been detected in this sample.  Using the flux
limits from our \textit{Spitzer} observations, we calculated the
fraction of the total 200,000 SED models from Robitaille et al.\
(2006) that we would have been able to detect.  Our calculated flux
limits were based on our three-sigma magnitude limits for the IRAC
bands for our observations: 14.3, 14.1, 12.4, and 11.6 mag for IRAC
3.6, 4.5, 5.8, and 8.0$\mu$m, respectively.  The SED models in the
grid are based on sources at 1 kpc, so the model fluxes were scaled to
4 kpc, the assumed distance of our sources.  The extinction to this
region ranges from 1 - 3 magnitudes in A$_{V}$ with an average
extinction of A$_{V}$ equal to 2 magnitudes (Moreno-Corral et al.\
2002 and Munari et al.\ 1998); this average A$_{V}$ was also accounted
for in the flux limits.  We found that with our flux limits we should
have been able to detect 42$\%$ of all models in the grid.  This
detection limit is based on our minimum flux limits in the high
background areas, and therefore applies to the fraction of objects we
should detect in the regions of higher nebular emission.
\begin{figure*}
\epsscale{1.0}
\plotone{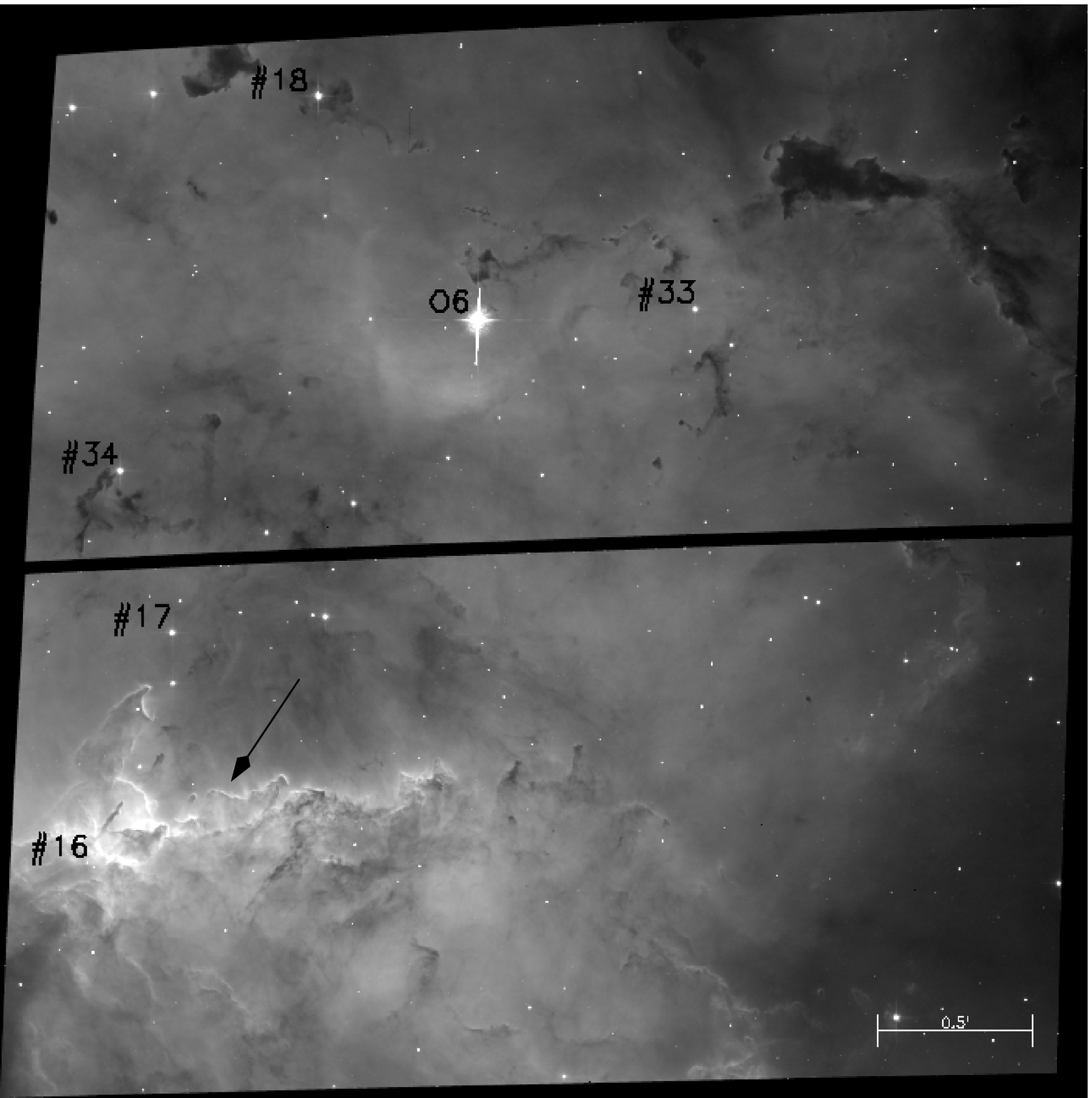}
\caption{\textit{HST} ACS F658N image of NGC 2467, centered on the main O6 V ionizing star. The \textit{HST} ACS image encompasses a total area of  4.2 $\times$ 4.2 pc, the scale bar on the image of 0\farcm5 corresponds to a distance of 0.6 pc. Detected YSOs in this image are labeled by their source number, there are three Class I/II sources, $\#$16, $\#$17, $\&$ $\#$18, and there are two Class II sources, $\#$33 $\&$ $\#$18. Four of the five YSOs shown here are located in close proximity to ionization fronts, and are seen sitting against dark clumps in the ACS image. Part of the main ionization front can be seen in this image and is identified by the arrow.}
\end{figure*}
\subsection{Comparison to \textit{HST} Images} 
The main motivation of our \textit{Spitzer} proposal was to observe a
select number of \hii~regions with \textit{Spitzer} that had already
been observed with \textit{HST}. \textit{HST} images provide us with
detailed information about YSOs after they are in an ionized
\hii~region environment, whereas \textit{Spitzer} allows us to see
both a larger view of the region and also to see protostars and their
disks that are still embedded in the dense gas around the \hii~region.

The \textit{HST} ACS images of NGC 2467 by De Marco et al.\ (2006)
were combined with our \textit{Spitzer} images in order to compare
what is seen with the different wavelength regime and better
resolution of \textit{HST} with the larger field of view of the
\textit{Spitzer} images.  The \textit{HST} field is centered around
the O6 V star, the brightest object in the \textit{HST} ACS image, and
a number of fragments, globules, and ionization front edges are also
seen in these images.  Five \textit{Spitzer}-detected YSOs are located
in the \textit{HST} field: three Class I/II sources and two Class II
sources.  Including the five YSOs, only a total of eight point sources
were detected in all four IRAC bands meeting our flux criteria set by
the high background emission and fall in the ACS FOV. There are,
however, more than eight point sources seen in the ACS image, as well
as in the two shorter wavelength IRAC bands in the ACS FOV. The
\textit{HST} ACS H$\alpha$ image of NGC 2467 is shown in Figure 6; the
five detected YSOs are labeled in the image by their YSO source
number; the O6 V star is also labeled.  Source $\#$16 is classified as
an IRAC Class I/II source and a MIPS Class I/O source, with a best-fit
mass of 4.7$\pm$0.3 M\sol~from our SED fitting. Source $\#$'s 17 and
18 are IRAC Class I/II sources with masses of 3.9$\pm$0.9 M\sol~and
4.3$\pm$0.5 M\sol, respectively.  Source $\#$'s 33 and 34 are IRAC
Class II sources with masses of 3.2$\pm$0.9 and 4.2$\pm$0.1 M\sol,
respectively.

Four out of the five YSOs ($\#$'s 16, 18, 33, and 34)  in the
\textit{HST} field are seen sitting against or near to dark clumps in
the ACS image.  There appears to be one main ionization front that 
\begin{figure*}
\epsscale{1.0}
\plotone{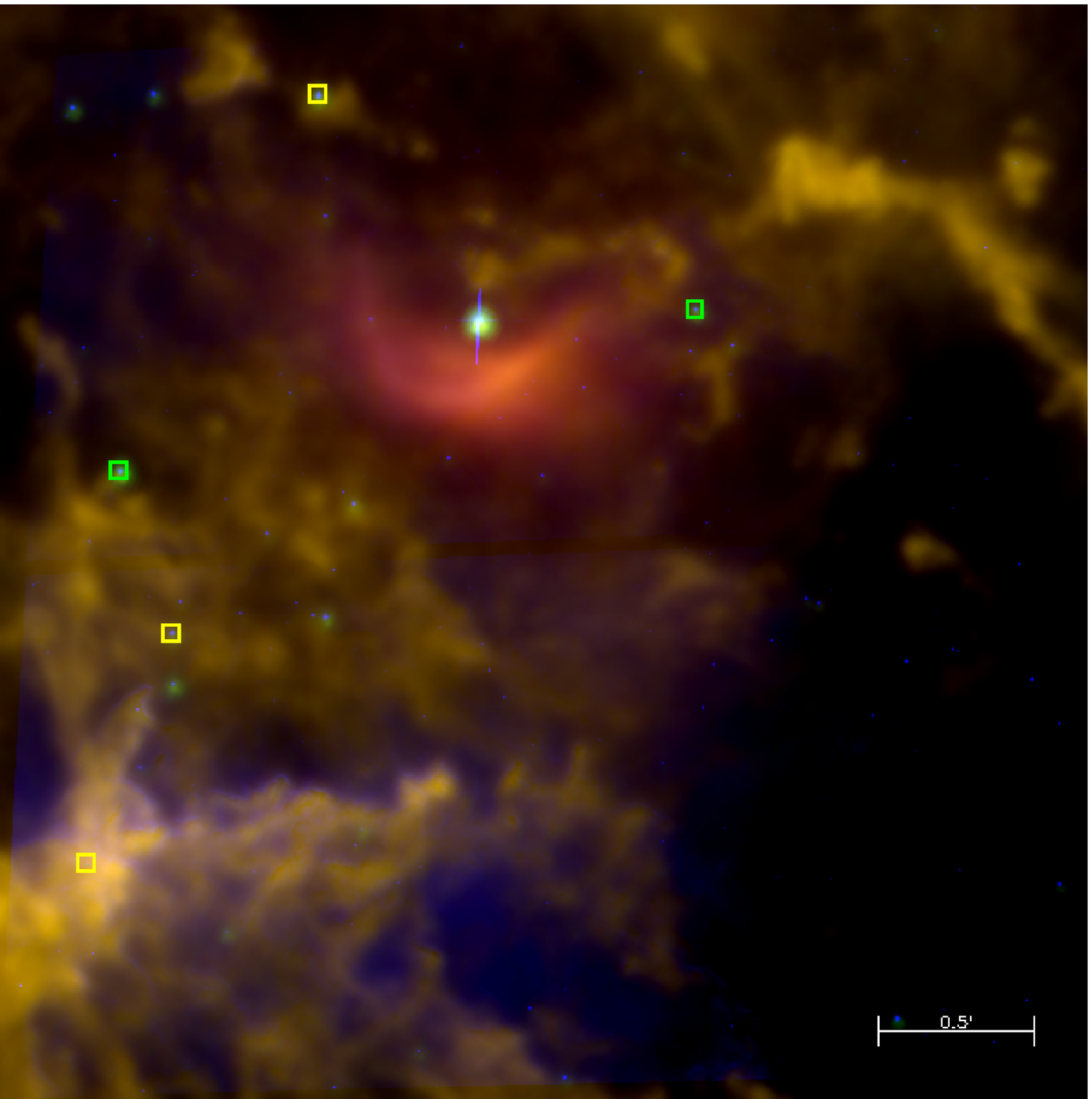}
\caption{3-color image of NGC 2467, scale of the \textit{HST} ACS image (total area equals 4.2 $\times$ 4.2 pc, with scale bar of 0\farcm5 equal to 0.6 pc). \textit{Spitzer} IRAC channel 4 is in red, IRAC channel 3 is in green, and ACS F658N is in blue. Detected YSOs in the image are marked by yellow and green squares, for the Class I/II and Class II YSO sources. The main ionization front, marked by the arrow in Figure 6, can clearly be seen here by the strong emission from the ACS F656N filter seen in blue.}
\end{figure*}
can be identified in the \textit{HST} image, located to the bottom left of
the image (see arrow in Figure 6).  This location corresponds to a
column of gas seen in the \textit{Spitzer} images where a large number
of YSOs are located. Figure 7 is a three-color \textit{HST} and
\textit{Spitzer} image of the region with the \textit{Spitzer} data
scaled to the \textit{HST} field of view.  The ionization front from
the main column in the \textit{Spitzer} bands is clearly outlined by
the ionization front as seen in the F658N filter. Three YSOs appear to
be in close proximity to this ionization front in the \textit{HST}
field. De Marco et al. (2006) make note of the many fragments that are
being uncovered by the advancing ionization front. Two of the
\textit{Spitzer} sources ($\#$'s 17 and 34) close to this ionization
front appear to already have been uncovered and are now sitting in the
interior of the \hii~ region. The third source located in close
proximity to the ionization front, $\#$16, is somewhat more confusing
as it appears to be embedded in one of these fragments, as viewed in
Figures 6 and 7.  The source is also somewhat hard to see in the ACS
image in Figures 6 and 7, but when viewed at full resolution in the
ACS image there is a possible faint identifiable point source
present. It also has a small best-fit extinction value (A$_{V}$ =
0.1). Therefore, it is plausible that source $\#$16 has also recently
been uncovered from the passing ionization front. One of the other
YSOs, $\#$18, at least in projection appears to be located in a denser
clump of material which can be seen in the F656N \textit{HST}
image. It does have a larger extinction value as best-fit from the SED
modeling with A$_{V}$ equal to 5.3, but the point source is clearly
visible in the ACS image. So, again it is hard to conclude if this
source is still embedded in the dense clump of material as seen in
Figure 7, and may likely have already been uncovered. The fifth YSO in
the \textit{HST} field is source $\#$33; it is close to the O6 V star,
and it also appears to be sitting inside the \hii~region.

One note of interest is that two of the Class I/II sources ($\#$'s 16
and 18) which are less evolved are possibly embedded sources, or are
at least seen sitting against dense clumps of material. In contrast,
the two Class II sources ($\#$'s 33 and 34) that are presumably more
evolved are located in the interior of the \hii~region, and have
already been uncovered by the ionization front from the O6 star.
Source $\#$17 is a Class I/II and it appears to also be located in the
interior of the \hii~region, but it is very close, within 0.2 pc, to
the edge of the ionization front and to a dense finger of gas. We
interpret this as meaning that it has recently been uncovered by the
ionization front.  This shows that even in the \textit{HST} field
alone, without detailed analysis, we already see a progression of
younger sources still embedded in the dense gas waiting to be
uncovered by the advancing ionization front, and older Class II type
sources sitting inside the ionized \hii~region having already been
uncovered by the ionization front.
  
\begin{figure*}
\epsscale{1.0}
\plotone{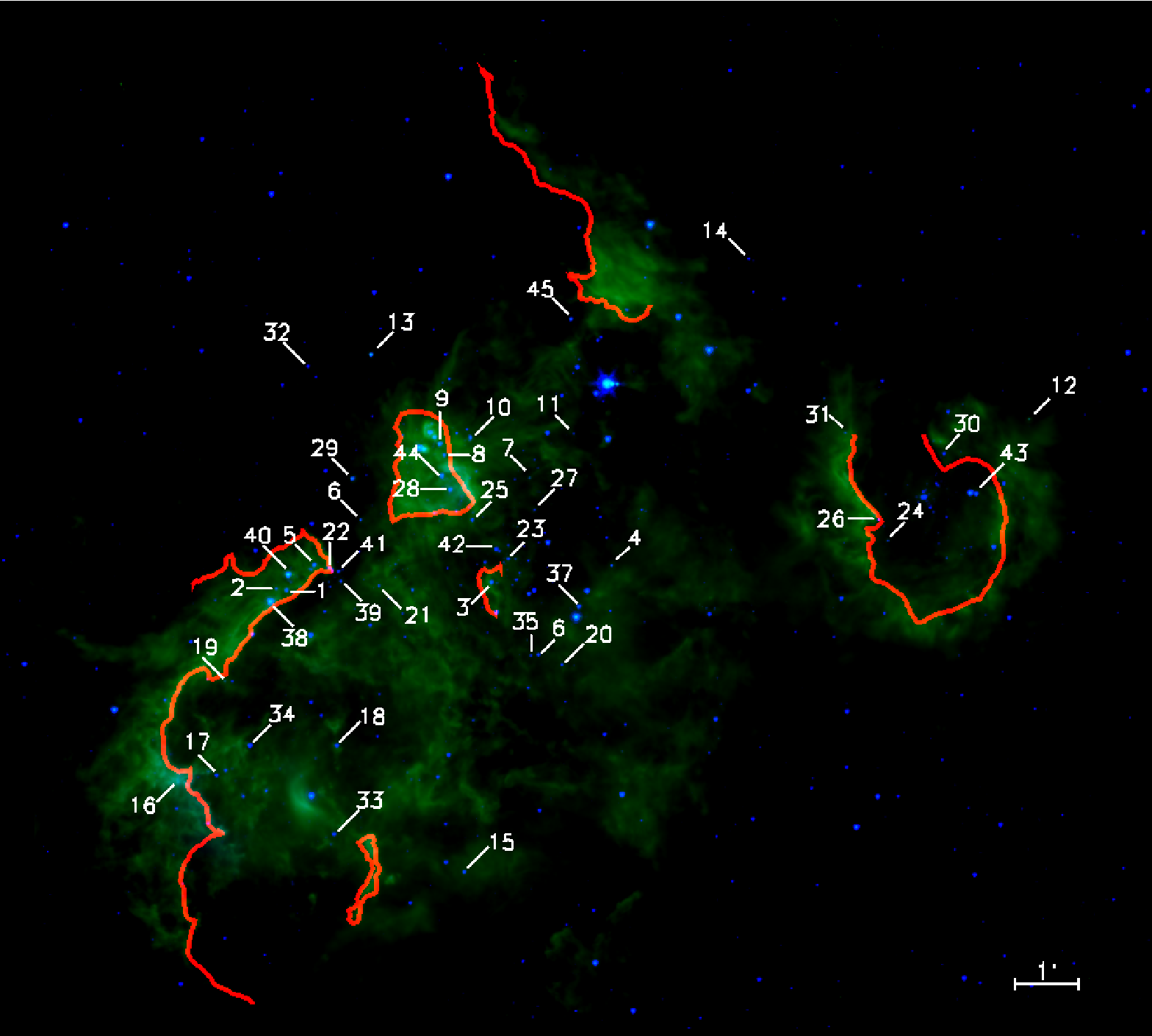}
\caption{Locations of identified ionization fronts in NGC 2467, same scale as in Figure 1.  Identified ionization fronts are outlined in red and are over-plotted on the IRAC 8.0 $\mu$m channel in green and the 4.5 $\mu$m channel in blue. The 45 candidate YSOs are labeled by their source number, and their location is shown. 29 out of the 45 YSOs appear to be strongly correlated with locations of identified ionization fronts. Line equal to 1\arcm~corresponds to a distance of 1.2 pc, given the assumed distance of 4.1 kpc to the region.}
\end{figure*}

\subsection{Spatial Distribution}

If there is no triggering occurring and the formation of most YSOs is
independent of the effects of the massive stars and \hii~region
expansion, then we would expect that the spatial distribution of YSOs
would not be correlated with compressed gas or ionization fronts. If
this is the case, then we would expect to see clusters of low-mass
stars already forming in the outer regions around \hii~regions.  On
the other hand, if there is a significant fraction of triggered star
formation in \hii~regions, then we would expect to see the
distribution of YSOs concentrated in the compressed gas and nearby to
ionization fronts (see figures 8 $\&$ 9 from Hester $\&$ Desch 2005
for a more detailed explanation of these scenarios and the
observational signatures of each).

We are interested in using our data to quantitatively distinguish
between these possible scenarios, triggered vs. non-triggered,
coeval star formation, or determine if there is a combination of
these modes occurring.  One possible way to do this is to look at the
distribution of protostellar sources in the region; if they are all coeval then we
would expect the YSOs to be distributed the same way as all other
sources, including the more massive OB stars. If triggering is occurring, 
we also want to estimate the mechanism of triggering from the expansion of the \hii~region.

As discussed in the introduction, there are a number of predictions for the three different proposed scenarios.  For all three models, one expects to see a correlation between the locations of the protostars and the ionization and shock fronts. For RDI, we expect no protostars to be forming in front of the ionization front, as it is the high pressure from the ionized gas that causes the clumps to implode.  For ``collect and collapse'', we expect to see regularly spaced protostars within the swept-up, compressed material.  For the third scenario, where triggering occurs from the shock front traveling in advance of the ionization front, one expects to see the youngest protostars in the compressed gas between the shock and ionization fronts, and to ages of protostars correlated with distance from the ionization front.  We tested these scenarios and triggered star formation vs.\ coeval star formation by
comparing the YSO distribution to the OB stellar population
distribution and to the location of ionization fronts in NGC 2467.

\subsubsection{Ionization Front Detection}
Ionization fronts were identified first coarsely, by examining the
images by eye, and then more precisely by running the edge detection
routine, \textit{roberts.pro} in astrolib of IDL, on the 5.8 and 8.0
$\mu$m band images. This routine performed gradient or directional
filtering, selecting locations where the pixel values changed by a
large amount from one pixel to the next in a given direction.  Six
areas in NGC 2467 had sharp edges visible in the smoothed image
produced by the edge detection routine and these 6 areas were
determined to contain possible ionization fronts. Once the six main
ionization fronts were identified, they were mapped in xy coordinates
using a routine in IDL that returns coordinate values; this routine
yielded an x and y position along the length of each ionization
front. The thickness of each ionization front outlined in the images
of NGC 2467 was assumed to be $\sim$10$^{17}$ cm thick, comparable to
typical thicknesses of ionization fronts (10$^{16}$ - 10$^{17}$ cm;
Osterbrock 1989). Figure 8 shows the locations of all 45 YSOs, labeled
by number, and the identified ionization fronts (on the same scale as
in Figure 1). The identified ionization fronts are outlined in red and
over-plotted on the IRAC 8.0 $\mu$m band in green and the IRAC 4.5
$\mu$ band in blue. 
\begin{figure}
\epsscale{1.2}
\plotone{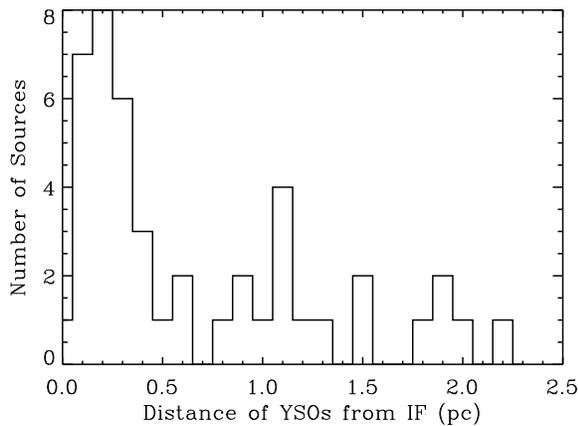}
\caption{Distance distribution of YSOs in NGC 2467 from the nearest ionization front, approximately 60$\%$ of the YSOs are located within a projected distance of 0.5 pc or less from the nearest ionization front.}
\end{figure}
The main ionization front identified is around the column on the left
of the image, as seen in Figures 1 and 8. The O6 V ionizing star is
near this column of gas; it has an average projected distance of 3 pc
away from the column.  Seven protostellar sources ($\#$'s 1, 2, 5, 16,
22, and 40) were identified in projection on this column, along with
three additional sources ($\#$'s 19, 39, and 41) which appear, in
projection, to be just past the ionization front presumably having
recently been uncovered.   All ten sources are very near (1-18\arcs)
to the edge of the ionization front.  Assuming a distance of 4.1 kpc
to NGC 2467, these 10 sources are located at projected distances
ranging from 0.02 to 0.35 pc from the edge of the ionization
front. The second region with an ionization front and a strong
clustering of sources is the central cluster in the middle of the
image.  This region is dominated by strong emission from a few B stars
in H18ab, the most massive is a B1 V star.  There are a large number
of detected YSOs near this sub-region, with the  eight closest
protostellar sources ($\#$'s 6, 8, 9, 10, 25, 28, 29, and 44) having
projected distances from this ionization front ranging from 0.12 to
0.73 pc.

The third most noticeable region is towards the upper right of the
image; this region is excited by the B1 V star in the H19 cluster.
The presence of a possible Str$\ddot{o}$mgren sphere around this cluster of
stars was revealed by narrow-band H$\alpha$, [N\,{\sc ii}], and
[S\,{\sc ii}] from Moreno-Corral et al.\ (2002). The ionization front
at the edge of this possible Str$\ddot{o}$mgren sphere is clearly defined in
the \textit{Spitzer} images of NGC 2467.  There are five protostellar
sources ($\#$'s 24, 26, 30, 31, and 43) sitting right near the edge of
the ionization front, with projected distances from the ionization
front ranging from 0.03 to 0.50 pc.  There is also one other source
($\#$12) slightly farther out, at a projected distance of 1.04 pc.

Three other ionization fronts were identified using the edge detection
routines.  Another five detected YSOs ($\#$'s 3, 23, 33, 42, and 45)
are located in close proximity to these three ionization fronts.
Overall, 29 of the 45 identified YSOs appear strongly clustered around
the detected ionization fronts, which  demonstrates that the majority
of the detected YSOs are most strongly concentrated around the
ionization fronts, and they are not well correlated with the O and B
star locations.

The histogram distribution of source distances shown in Figure 9 shows
that there is a markedly higher frequency of sources near the
ionization fronts.  The projected distances of the 45 YSOs from the
nearest ionization front ranges from 0.02 - 2.2 pc, with more than
60$\%$ (28 out of 45), of the sources falling within 0.6 pc of a
detected ionization front, and 70$\%$ (32 out of 45 objects) falling
within 1.0 pc of an ionization front.  Our analysis is incapable of
detecting face-on ionization fronts; therefore the other 13 sources,
which are not seen in projection within 1.0 pc of an ionization front,
may still be near ionization fronts as well.

\subsubsection{Distribution Tests}
The average projected distance of the edge-on ionization fronts in NGC
2467 is about 3 - 5 pc away from the OB stars. A simplified
3D-model\footnote[1]{A cylindrical bowl, with equal height and width
was used to model the distribution.} of a hemispherical ionization
front which is on average 3 to 5 pc away from an O star was compared
to the distribution of sources we see in NGC 2467.  This model was
used in order to determine if it is likely that the other YSOs not
seen in projection within 1 pc from an ionization front are in fact
closer to the ionization front than they actually appear to be. We
were only able to detect parts of the ionization fronts that are
moving edge on to our line of sight.  However, we would not be able to
detect the part of the ionization front that is face-on. We therefore
want to determine where this total predicted distribution of sources
matches the observed distribution of YSOs in NGC 2467.

In order to calculate the predicted distribution of sources within a
layer a given distance away from the ionization front, we calculated
the ratio of the edge-on volume of this distribution (which we would
be able to observe) to the total volume of this distribution. The
thickness of the expected YSO distribution was measured from 0.1 to
2.2 pc. As the thickness of the distribution increased, the larger the
percentage of expected observed sources within that distribution layer
becomes.  Varying the radius from 3 to 5 pc (the distance of
ionization front to the OB star) only slightly changes the predicted
fraction of sources at a given thickness. Using this model, we
calculated that a distribution that is concentrated within a layer 0.5
pc away from an ionization front would show 68$\%$ of the sources
within a projected distance of 0.5 pc from this layer when viewed in
projection.  For a distribution that is concentrated within 1 pc, we
calculated that 70$\%$ or less of the sources should be seen in
projection within 1 pc. However, for the actual distribution of YSOs
in NGC 2467 almost 60$\%$ of them are seen in projection within 0.5 pc
of an ionization front and over 70$\%$ are seen within 1 pc of an
ionization front.  
\begin{figure}
\epsscale{1.15}
\plotone{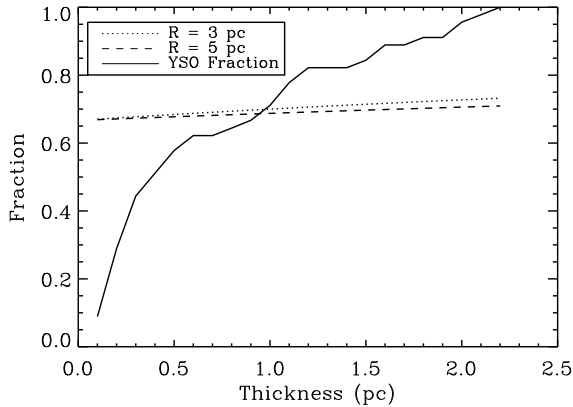}
\caption{A simplified 3D hemispherical model of an ionization front is used to determine the expected number of observed sources that would be seen within the layer thickness when viewed in projection.  Fraction of expected sources observed within a given thickness distribution from an ionization front, with ionization fronts centered at 3 and 5 pc away from an OB star shown by dashed lines. Actual distribution of YSOs in NGC 2467 shown by solid line. The expected distribution and the observed distribution cross at approximately 1 pc, indicating that the actual distribution of YSOs in NGC 2467 is a population of sources concentrated within 1 pc or less from the nearest ionization front.}
\end{figure}

\begin{figure}
\epsscale{1.15}
\plotone{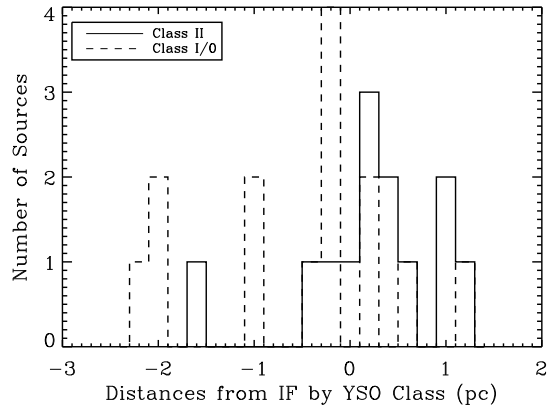}
\caption{A histogram plot of the distances of Class II and Class I/0 YSOs from the nearest ionization front. Negative distances represent YSOs ahead of the ionization front, still embedded in compressed gas, positive distances for sources behind the ionization front that have already been overrun by the passing front.  A separation between the two classes of protostellar objects can be seen, with a majority of the youngest YSOs (Class I/0) being located ahead of the ionization front in compressed gas, and a majority of the oldest objects (Class II) located behind the front.}
\end{figure}
Figure 10 shows the predicted fractions of sources
for the 3D-hemispherical model centered at 3 and 5 pc versus
thickness, along with the actual fraction of YSOs that are within the
same distance (thickness) from an ionization front.  The observed YSO
distribution crosses the predicted model fractions at a distance
between 0.9 and 1.0 pc. Above a distance of 1 pc, the YSO fraction is
greater than the expected distribution.  This suggests that the
overall distribution of YSOs in NGC 2467 is a population of sources
that is concentrated within a layer that is 1.0 pc or less away from
an ionization front.

A second statistical method was used in order to test the likelihood
of this current distribution occurring.  We calculated the probability
of the YSO sources being randomly distributed versus the actual
observed distribution. We calculated the fraction of the survey area
that is within a given distance of the ionization fronts.  If this is
a random distribution then we would expect the fraction of YSOs
located within a given distance of an ionization front to equal the
fraction of the total survey that is within that distance. For
example, only 4$\%$ of the survey area is within 0.2 pc of an
ionization front, but 10 out of 45 YSOs (22$\%$) are located within
this distance.  The probability of 22$\%$ of all randomly distributed
YSOs falling within 0.2 pc of the ionization front by chance is only
0.001$\%$.  This was calculated using a Poisson probability
distribution, which gives the probability of a specific number of YSOs
(k) being distributed randomly, as shown by Equation 3:

\begin{equation}
{Prob = \frac{\lambda^{k} e^{-\lambda}}{k!}}.
\end{equation}

The expected number of sources ($\lambda$) is equal to the fraction of
the survey area within a given distance multiplied by the total number
of sources (45) and k is equal to the observed number of YSOs within
that distance. The probability of a random distribution drops to as
low as 10$^{-12}$ at a distance of 0.5 pc from the ionization front,
where 26 out of 45 YSOs are located within this distance, despite only
11$\%$ of the total survey area being within this distance.  Looking
at the likelihood of all 45 YSOs being located within a distance of
2.2 pc of the ionization fronts, we find the odds of this occurring by
chance to be 10$^{-6}$.  Therefore, we conclude that this is not a
random distribution of sources.  The data indicate that the locations
of YSOs are strongly correlated with the location of the ionization
fronts.

\subsubsection{Distance and Age with Evolutionary Class}
We also examined the distribution of YSOs for each specific
evolutionary class in order to determine if there is any noticeable
trend in distance from ionization fronts versus evolutionary class,
and to also help in determining how the triggering might be occurring.
If RDI is occurring, we expect no protostellar objects to be located
ahead of the ionization front because in this scenario the high
pressure ionized gas leads to the implosion of pre-existing clumps.
For the other two scenarios, the shock front preceding the ionization
front triggered the collapse.  The youngest protostars will be located
in the compressed gas between the two fronts, and there should be an
age progression with distance from the ionization front.

For this analysis, we looked at the distance of each YSO from the
ionization front in terms of those already uncovered versus those
still embedded.  YSOs that appear to be ahead of the ionization front
and are still embedded are considered to have a negative distance from
the ionization front.  YSOs that are behind the passing ionization
front and have already been uncovered are given a positive distance.
Figure 11 shows a histogram of the distance of the Class II and Class
I/0 YSOs from the nearest ionization front, with positive distances
for YSOs behind the front and negative distances for those ahead of
the ionization front.  In this plot it is fairly clear to see a
separation between the two different types of YSOs.  The majority of
the Class II YSOs have already been overrun by the passing ionization
front (10 out of 13 with positive distances), and on the other hand a
majority of the Class I/0 YSOs (10 out of 14 with negative distances)
are still embedded in compressed molecular gas and have yet to be
uncovered by the ionization front.  The Class I/II YSOs, which are not
plotted in Figure 11, fall in between these two distributions, with
two-thirds (12 out of 18) located behind the front, and the other
third still embedded in compressed gas ahead of the ionization
front. This suggests that we see a trend of evolution with distance
from an ionization front.

We can also rule out RDI as a dominant mechanism for triggered star
formation in this region because we do see a number of protostellar
objects located ahead of the ionization front, suggesting that the
YSOs are being triggered from the shock front traveling in advance of
the ionization front.  We favor the shock front traveling in advance
of the ionization front as the triggering mechanism over the ``collect
and collapse'' scenario because we do not see strong evidence for
regularly spaced protostars forming around the edge of the
\hii~region.  We also see strong evidence, as shown in Figure 11, for
a distribution of YSOs with ages that correlate with distance from the
ionization front.  However, it is still plausible that some YSOs in
this region formed by the ``collect and collapse'' mechanism as well.
The third triggered star formation scenario can also account for why
we see a lack of protostars close to the OB stars, ie.) between the O
star and the ionization fronts.  This gap exists because we cannot
directly observe stars older than a certain age with this dataset,
those that would be located nearest the OB stars.  Either no other
stars formed near the OB stars, which does not seem plausible, or
stars formed near the OB stars first and all the new protostars
forming now are spatially correlated with the ionization fronts.
Therefore the lack of new protostars currently near the OB stars, and
the correlation of new protostars with the ionization fronts, over wide
spatial scales in the region, is direct evidence for a triggering
mechanism.

An argument against the triggered star formation scenario is that many
of these objects may have been forming before the shock front passed
over them. In this case, they are just being uncovered by the passing
ionization front, and not triggered by the compression of the
molecular gas due to the expanding shock front.  This therefore partly
becomes an age argument; if the majority of the sources have ages much
greater than the timescale between the passage of the shock front and
the ionization front, then it would be possible that they were forming
before the shock front compressed the surrounding molecular gas.  The
passage of the shock front is typically followed by the passage of the
ionization front within a few$\times$10$^{5}$ yr (Hester and Desch
2005; Sugitani et al.\ 2002).

For the 32 sources that fall within a projected distance of 1 pc or
less from an ionization front, $\sim$70$\%$ have best-fit ages less
than or equal to 10$^{5}$ yr and $\sim$80$\%$ have ages less than
3$\times$10$^{5}$ yr. For all 45 YSOs, $\sim$65$\%$ have ages less
than or comparable to the timescale between the passage of the shock
front and the ionization front.   Although the SED model best-fit ages
may not be completely reliable, we can also use the calculated
distances of the YSOs from the ionization fronts and can compare the
time elapsed since the ionization front passed the YSOs (or will pass
them) to their lifetimes.  Assuming an ionization front speed of 0.5 -
2 km s$^{-1}$ (Osterbrock 1989), sources at a projected distance of
0.5 pc from an ionization front are only 2.5$\times$ 10$^{5}$ -
1$\times$10$^{6}$ yr away from the nearest ionization front. YSOs that
are closer than 0.5 pc were passed by the ionization front even more
recently (or will be uncovered by the advancing ionization front even
sooner).  These timescales are comparable to the best-fit ages and
expected lifetimes of Class 0, Class I, and Class II protostellar
objects. We conclude that we are seeing very young objects that are
still forming, many that are even younger than 10$^{5}$ yr. This
demonstrates that when the majority of the YSOs formed in the last
few$\times$10$^{5}$ yr they likely formed in gas compressed by the
shock front from the expanding \hii~region, and are now forming very
near to ionization fronts.

The most notable thing we see with the distribution of these sources
is that they are strongly associated with the ionization fronts, and
they are not randomly distributed. We also do not see clustering of
YSOs around the OB stars, which seems to suggest that they are not
correlated with the distribution of the OB stars themselves, but are
instead more correlated with the ionization front locations. Also, we
see no discernible separation in the distribution of the low-mass and
the intermediate mass YSOs from the ionization fronts.  It appears
that low-mass sources are just as likely to be triggered by the
expansion of the \hii~region as the more massive YSOs.  Therefore it appears that the star formation is not strictly coeval in NGC 2467,
and that triggered star formation due to the expansion of the
\hii~region is occurring. Overall, the distribution of the candidate
YSOs in this region suggests that a clear majority of the current
protostars are forming in regions of gas that are being compressed
from the advancing shock and ionization fronts.

\subsection{Estimates of Triggering and Star Formation Rates}

A final way to test the various scenarios for star formation in this region
is to estimate the amount of triggering that has occurred in NGC 2467
throughout the lifetime of the region. First, we estimate the total
star formation rate (SFR) in NGC 2467. Using the Salpeter IMF, we
determined that there should be a total of $\sim$5000 stellar sources
ranging in mass from 0.2-40 M\sol (see equation 2). We calculate the
total mass in stars from the Salpeter mass distribution (Salpeter
1955), giving a total of 3500 M\sol, as shown in Equation 4.  From
there, we calculate the average SFR in NGC 2467 over the last 2 Myr,
given by Equation 5:

\begin{eqnarray}
M_{tot} &=& \int_{0}^{N} M dN = A\int_{M_{1}}^{M_{2}} M^{-1.35} dM \nonumber \\
&=& 8\times10^{2}\int_{0.2M_{\odot}}^{40M_{\odot}} M^{-1.35} dM \sim 3500 \,{\rm M_\odot}
\end{eqnarray}

\begin{equation}
Average~SFR = \frac{3500 \,{\rm M_{\odot}}}{2 Myr} = 1.75\times10^{-3}~ \,{\rm M_{\odot}} \,{\rm yr}^{-1} 
\end{equation} 

In \S4.2, we showed that from our \textit{Spitzer} flux limits we
should be able to detect 42$\%$ of all YSO sources based on the grid
of SED models from Robitaille et al.\ (2006), however this estimate
disregarded the age of the sources and age of the region.  Therefore,
in order to calculate the total triggered star formation rate, we want
to measure how the detection ability of individual sources would
change with elapsed age in the region and age of the individual
source.  We want to include only realistic models, because the
fraction of detectable sources depends on the number of models we
compare to.  To do this, we took a population of stars given by the
Salpeter IMF, between our mass limits of 0.2 and 40 M\sol, and assumed
a constant SFR equal to the average SFR from Equation 5
(1.75$\times$10$^{-3}$ M\sol~yr$^{-1}$), and then used the Salpeter
IMF to age that population of stars.  We looked at the detection
probabilities for this population of stars at timesteps every 10$^{5}$
yr.  During each timestep we would take the population of stars at
that specific age and of various masses (determined by the IMF), and
determine the percentage of sources, from the SED models, we should
detect based on our \textit{Spitzer} flux limits.  The population of
stars was aged from one timestep to the next, with a new population
of young stars added in at each timestep.
\begin{figure}
\epsscale{1.2}
\plotone{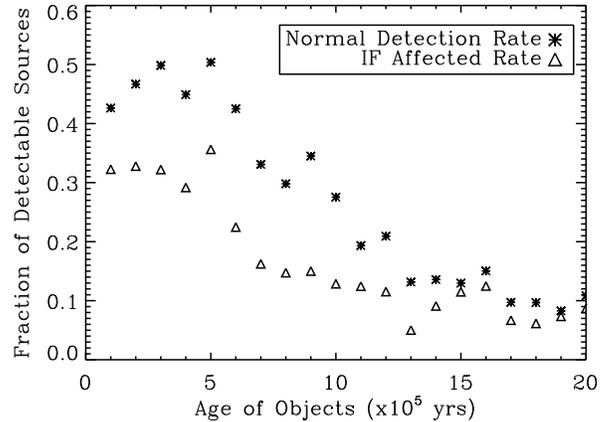}
\caption{The fraction of detectable sources vs.\ age of the object. A population of stars given by a Salpeter IMF and a constant SFR (value equal to the calculated average SFR in NGC 2467) was aged over timesteps of 10$^{5}$ yr. The fraction of detectable sources based on our flux limits was calculated at each timestep, the normal detection rate is shown by the asterisk symbols. The detection rate was also calculated for sources uncovered by a passing ionization front. Disk erosion was assumed using models from Johnstone et al.\ (1998). The ionization front affected detection rate with object age is shown by the triangles. The data show that the best age range to detect the YSOs is around a few$\times$10$^{5}$ yr.}
\end{figure}

\begin{figure}
\epsscale{1.2}
\plotone{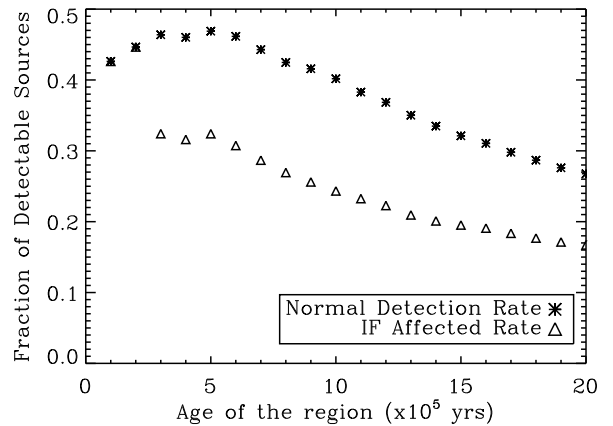}
\caption{The fraction of detectable sources vs.\ age of the region, using the same methods as described for Figure 12.  The normal detection fraction is shown by the asterisk symbols and the ionization affected fractions are shown by the triangles. The detection rate drops to below 20$\%$ by 2 Myr.}
\end{figure}

In Figure 12 we show the fraction of detectable objects versus age
of the individual sources.  In this plot, we see that the best range
of ages to detect the YSOs is from 0.1 to 0.5 Myr, and the peak age is
at 3$\times$10$^{5}$ yr.  After 0.5 Myr, the detection rate drops off
dramatically.  Within the first 0.5 Myr, the fraction of detected
sources is near 50$\%$, but by the time the YSOs reach an age of 2
Myr, the fraction that would be detected has fallen to less than
15$\%$.  This implies that the YSOs we are detecting likely have ages
of a few$\times$10$^{5}$ yr, agreeing with best-fit ages from the SED
fitter and timescales from the ionization fronts.

In Figure 13 we have plotted detection probability vs. elapsed age of
the region, with probabilities being calculated every 10$^{5}$
yr. This plot shows the detection rate of YSOs integrated over the
star formation history of the region. During the first 0.5 Myr the
detection rate was almost 50$\%$, which is comparable to the total
completeness estimates of 42$\%$ that we found in \S4.2; but as
the age of the region increases and the population of stars ages, our
detection rate decreased to only 27$\%$ by 2 Myr.

One other factor that could decrease our detection ability is the fact
that once the ionization front overruns the newly forming protostar it
will quickly begin to erode the protostellar disk. Johnstone,
Hollenbach and Bally (1998) calculated that the disk mass loss rate
for low-mass stars due to nearby massive stars is $\sim$10$^{-7}$
M\sol~yr$^{-1}$.  St$\ddot{o}$rzer and Hollenbach (1999) show that
for sources in Orion at distances greater than 0.3 pc from the
ionizing source, the disk mass loss rate is dependent on the initial
size of the disk and the distance from the ionizing source. A recent study of disk
survival in NGC 2244 using \textit{Spitzer} data (Balog et al.\ 2007),
has also shown that it is possible that effects from high-mass stars on
disk survival is only limited to distances of 0.5 pc or less from the
high-mass stars. Therefore, this disk mass loss rate
may not be as strong in NGC 2467 as it was modeled for stars in the
Trapezium Cluster; however, we are still interested in how disk
photoevaporation can effect our detection abilities of YSOs located
within close proximity to the ionizing sources in this region. This
effect may be one reason why we do not detect more protostars
located at very nearby distances to the massive stars.  

Using the same
method as above, with the weighted grid of models according to the
calculated IMF and SFR, we looked at what fraction of sources would
still be detected if the disk mass was decreased after the source was
uncovered by an ionization front.  Again, we want to consider
realistic models, therefore we consider only models which have an
initial disk mass of 10$^{-8}$ M\sol~or greater. Including all models
from the YSO SED grid would lower the total fraction of detectable
sources. However, by setting a minimum disk mass and a minimum mass of
the object (0.2 M\sol) we do not bias ourselves towards a lower,
unrealistic fraction of detectable sources. The effect of different
model disk masses on the detection calculations was tested by varying
the lower limit of the initial disk mass from 10$^{-9}$ - 10$^{-7}$
M\sol.  We found that the detection fractions only changed by a few
percent between these different initial disk mass values, therefore
the average value of 10$^{-8}$ M\sol~for the minimum disk mass was
used. Photoevaporation of the ``EGG'' and proplyd phases will be
relatively short-lived stage, $\sim$10$^{4}$ yr during each phase
(Hester $\&$ Desch 2005), ending with disks typically eroded down to
sizes of $\sim$30 - 50 AU. Photoevaporation of the rest of the
truncated disk will take several million years (Johnstone et al.\
1998). We therefore  assume, for simplicity, that the majority of the
photoevaporation takes place on timescales of few$\times$10$^{4}$ yr
(Hester $\&$ Desch 2005).  Under these assumptions, we found that the detection rate for sources in close proximity to the ionizing source could drop from the initial 45$\%$ to as low as 17$\%$, and again the best age range to detect these objects is around a few$\times$10$^{5}$ yr.  These data are also shown in Figures 12 and 13.

Another note of interest, as seen in Figures 12 and 13, is that while
we assumed a constant star formation rate over the lifetime of the
region the detection rate decreases with time.  Haisch et al.\
(2001) showed that in young star forming clusters, the fraction of
sources with disks decreases with age of the cluster.  The results
presented here seem to suggest that even though the same number of
stars are being formed at each timestep, the total detection fraction
is still decreasing, similar to the decrease in disk fraction measured
by Haisch et al.\ (2001).  Other recent studies have used the Haisch
et al.\ result to justify that the star formation rate is dramatically
decreasing with cluster age (Gounelle $\&$ Meibom 2008). However, it
seems plausible that the star formation rates in young clusters may
not be decreasing as dramatically after the first few Myr as claimed
by Gounelle $\&$ Meibom (2008), only that the detectability of the
total fraction of stars with disks decreases with time as demonstrated
by Figures 12 and 13.  

As shown in \S4.4, we currently observe 45 YSOs, the majority of
which have been influenced by the effects of \hii~region
expansion. In order to look at the overall rate of triggered 
star formation, we need to estimate the average age of the current
population of YSOs. From the SED model fitting, the 45 YSOs have
best-fit ages that average to $\sim$2$\times$10$^{5}$ yr. From the
previous analysis, we show that best age to detect the YSOs is
$\sim$3$\times$10$^{5}$ yr. In \S4.4.3, we also demonstrate that
the due to the close proximity of a majority of the YSOs to the
ionization fronts, the timescales are likely to be only a
few$\times$10$^{5}$ yr.

Multiple methods have shown that the 45 current YSOs likely have
average ages of few$\times$10$^{5}$ yr. While the actual detection
fraction value might be somewhat uncertain, it is likely that it is at
least below the 30$\%$ level, and under plausible assumptions for sources in close proximity (0.5 pc or less) to the ionizing OB stars, one might imagine that the
detection fraction may be as low as 17$\%$.  If our ability to detect
the current population of sources is 17 - 27$\%$ compared to what it
would have been during the first 0.5 Myr in the region, then this
would also mean that the actual total number of current YSOs could be
as much as five times larger. Therefore as shown in Equation 6, 45
detected YSOs with an average age of 2 - 3 $\times$10$^{5}$ yr results
in 6 - 13$\times$10$^{-4}$ stars formed per yr that are triggered.

\begin{eqnarray}
\frac{\#~YSOs}{Avg. Age \times Detection Rate} &=&  \frac{45}{(2-3\times10^{5}~ \,{\rm yr}) \times (0.17-0.27)}  \nonumber \\
 &=& 13 - 6\times10^{-4}~\,{\rm stars} \,{\rm yr}^{-1} 
\end{eqnarray} 
Using the Salpeter IMF this results in a triggered SFR of $\sim$ 4.2 -
9.0$\times$10$^{-4}$ M\sol~yr$^{-1}$. If this rate of triggered star formation has been constant over the
age of the region then by comparing this rate to the total SFR, given in Equation 6, this results in 24 - 52$\%$ of the total SFR.  We therefore conclude that approximately 25 - 50$\%$ of
the YSOs in NGC 2467 may have been formed due to triggering from
\hii~region expansion. 

There are a number of assumptions that have gone into this
calculation: assuming a constant SFR, an estimate of the average age
of the YSOs from multiple methods, and the determination of the
current detection fraction of YSOs.  Our assumption about the
detection fraction of YSOs nearby the massive stars may be one factor
that could change our triggered SFR estimate. If the current detection
fraction is not adversely affected by the photoevaporation of YSOs
nearby the massive stars, or if most of the current population is not
close enough to a massive star, then the detection fraction may only
be as low as 27$\%$.  This would result in an estimated triggered star
rate of 25-33$\%$ of the total SFR, and while this lowers our
calculated triggered SFR it still demonstrates that some fraction of
the YSOs in this region are likely to be triggered.  While admittedly
the triggered star formation calculation is a somewhat simplified
calculation, it does allow us to obtain some estimate for how much
triggering is occurring in this region.  Both the total SFR and the
triggered SFR are average values over the last 2 Myr, but we would
expect the entire process for triggered star formation to have been
higher earlier on due to densities being higher nearest to the O
stars, and shocks moving at faster rates when the density gradient is
higher. Therefore, the estimates of 25 - 50$\%$ may represent lower
limits for the fraction of stars that would have been affected by
triggering from \hii~region expansion. Although it is probable that
not all of the YSOs in NGC 2467 have formed from triggering
mechanisms, the evidence seems to suggest that some amount of
triggering is occurring due to the expansion of the \hii~region. Even
as a lower limit, 25 - 50$\%$ is a significant fraction of the total
amount of star formation occurring in NGC 2467; therefore, this cannot
be discarded as a possible mode of star formation.

\section{Summary and Conclusions}
The mid-IR data from \textit{Spitzer} show that NGC 2467 is a
region of active star formation.  We detected a large number of
sources with mid-IR excesses, which is evidence of the youth of the
region. 45 YSOs were detected based on their \textit{Spitzer}
colors, and 27 of them also showed an infrared excess in their 2MASS
colors. 

One of the main questions about star formation that we wanted to
address is whether or not there is evidence for triggered star
formation or if the population is coeval.  Triggered star formation by
\hii~region expansion appears to be occurring in NGC 2467.  The
current population of protostars is spatially correlated with the
ionization fronts.  The new protostars appear to be forming in a
coherent structure which is associated with the ionization fronts.  We
have also shown, based on SED fitting, that this population is very
young, with the majority having ages less than a few$\times$ 10$^{5}$
yr, much younger than the age of the region and of the ages of the OB
stars.  We therefore conclude that the star formation in NGC 2467 has
been ongoing, and it is not coeval, with strong evidence for
triggering from the expansion of the \hii~region created by the
massive OB stars.

We favor an ionization front with a preceding shock as the most likely
triggering mechanism, but whichever mechanism, we can estimate the
amount of triggered star formation that is occurring.  The estimated
current total star formation rate in NGC 2467 is 1.75$\times$10$^{-3}$
M\sol~yr$^{-1}$, with a triggered star formation rate equal to 0.42 -
0.90$\times$10$^{-3}$ M\sol~yr$^{-1}$. Based on the distribution and
age estimates of the current population of candidate YSOs, we estimate
that between 25 - 50$\%$ of the YSOs forming in this region are due to
triggering from the advance of ionization fronts created by nearby
massive stars.  It is probable that some of the low-mass stars in NGC
2467 have formed spontaneously throughout the lifetime of the region,
but we are also seeing a population of very young objects that are
currently forming in compressed gas from the \hii~region
expansion. While the data do not suggest that triggering is the only
mode of low-mass star formation in this region, it does suggest that
triggering is occurring in NGC 2467 and has contributed a substantial
fraction to the overall star formation rate.

This work has implications for other star forming regions and even for
possible mechanisms of our own Sun's formation.  Although this is only
one region, we can attempt to use the results from this region to
predict what fraction of all low-mass stars may be triggered by
\hii~region expansion.  Lada and Lada (2003) have shown that as much
as 70-90$\%$ of all low-mass stars form in rich clusters also
containing massive stars, therefore if 25-50$\%$ of low-mass stars in
rich clusters are triggered, then as many as 18-45$\%$ of all low-mass
stars may be formed by this mechanism. Possibly including our own Sun
and Solar System, where meteoric evidence for live $^{60}$Fe, a
short-lived radionucleide, in the early solar system suggests that a
supernova event occurred shortly before the formation of the planets
(Tachibana $\&$ Huss 2003; Tachibana et al.\ 2006).

The environment created by the massive stars in NGC 2467 is having
a definite impact on the formation of further generations of stars.
We find that newly forming YSOs are mostly found in areas
where the shock front driven in advance of the ionization front is
compressing the molecular gas.  The sources we are detecting with
mid-IR excesses are very young sources, with the majority having ages
less than a few $\times$ 10$^{5}$ yr. Thus the sources we are finding are 
not coeval with the OB stars. The 45 YSOs detected and the
distribution of these sources suggest that NGC 2467 is a prime example
of triggered star formation due to \hii~region expansion.

\acknowledgements  
We wish to thank Angela Cotera and Steven Finkelstein
for many helpful conversations. We also wish to thank the anonymous referee for useful comments and suggestions that greatly improved this paper.
This work is based in part on observations made with
the \textit{Spitzer Space Telescope}, which is operated by the Jet
Propulsion Laboratory, California Institute of Technology under a
contract with NASA. Support for this work was provided by NASA through
an award issued by JPL/Caltech, program PID 20726. This work was also
supported by the Arizona State University (ASU) Department of
Physics, the ASU School of Earth and Space Exploration and the
ASU/NASA Space Grant.

\end{document}